%% file: TurbulentSF_v12.tex
\documentclass[referee,usenatbib]{mn2e}
\usepackage{epsf}
\usepackage{graphicx}

\input{journals}

\newcommand{\sm}[1]{\mbox{{\scriptsize #1}}}

\newcommand{\simge} {\,{}^>_{\sim}\,}

\newcommand{\be}{\begin{equation}}
\newcommand{\ee}{\end{equation}}
\newcommand{\bea}{\begin{eqnarray}}
\newcommand{\eea}{\end{eqnarray}}
\newcommand{\bdm}{\begin{displaymath}}
\newcommand{\edm}{\end{displaymath}}
\newcommand{\bef}{\begin{figure}}
\newcommand{\eef}{\end{figure}}
\newcommand{\befone}{
  \begin{figure*}
  \centering
  \begin{minipage}{\textwidth}
  }
\newcommand{\eefone}{\end{minipage}\end{figure*}}

\newcommand{\refeq}[1]{(\ref{#1})}

\newcommand{\cm}{\mbox{cm}}

\newcommand{\km}{\mbox{km}}
\newcommand{\AU}{\mbox{AU}}
\newcommand{\ergs}{\mbox{ergs}}
\newcommand{\g}{\mbox{g}}

\newcommand{\Msol}{\mbox{$M_{\sun}$}}
\newcommand{\pc}{\mbox{pc}}

\newcommand{\K}{\mbox{K}}
\newcommand{\yr}{\mbox{yr}}
\newcommand{\ys}{\mbox{yrs}}
\newcommand{\Myr}{\mbox{Myr}}

\newcommand{\Ma}{{\cal M}}

\newcommand{\di}{\mbox{d}}

\newcommand{\Htwo}{\mbox{H$_2$}}

\def\eps@scaling{0.98}

\def\showone#1{
  \centering
  \leavevmode
  \epsfxsize=\eps@scaling\linewidth
  \epsfbox{#1.eps}
}

\def\epstwo@scaling{0.48}

\def\showtwo#1#2{
  \centering
  \leavevmode
  \epsfxsize=\epstwo@scaling\linewidth
  \epsfbox{#1.eps} 
  \epsfxsize=\epstwo@scaling\linewidth
  \epsfbox{#2.eps}
}

\def\showthree#1#2#3{
  \centering
  \leavevmode
  \epsfxsize=\eps@scaling\linewidth
  \epsfbox{#1.eps} 
  \epsfxsize=\eps@scaling\linewidth
  \epsfbox{#2.eps}
  \epsfxsize=\eps@scaling\linewidth
  \epsfbox{#3.eps}
}

\def\showsix#1#2#3#4#5#6{
  \centering
  \leavevmode
  \epsfxsize=\epstwo@scaling\linewidth
  \epsfbox{#1.eps} \hfil
  \epsfxsize=\epstwo@scaling\linewidth
  \epsfbox{#2.eps} \hfil
  \epsfxsize=\epstwo@scaling\linewidth
  \epsfbox{#3.eps} \hfil
  \epsfxsize=\epstwo@scaling\linewidth
  \epsfbox{#4.eps} \hfil
  \epsfxsize=\epstwo@scaling\linewidth
  \epsfbox{#5.eps} \hfil
  \epsfxsize=\epstwo@scaling\linewidth
  \epsfbox{#6.eps}
}

\title[Filamentary accretion and massive star formation]
      {Supersonic turbulence, filamentary accretion,      
	and the rapid assembly of massive stars and disks}
      
\author[R.~Banerjee, R.E.~Pudritz \& D.W.~Anderson]
       {Robi Banerjee$^{1,2}$, Ralph E. Pudritz$^{2,3}$ and Dave
       W. Anderson$^2$ \\
         $^1$Institute of Theoretical Astrophysics (ITA), University of
       Heidelberg, Germany \\
	 $^2$Department of Physics and Astronomy, McMaster University,
       Hamilton, 
	 Ontario L8S 4M1, Canada \\
	 $^3$Origins Institute, McMaster University, Arthur Bourns Bldg 241,
	 Hamilton, Ontario L8S 4M1, Canada}

\begin{document}

\maketitle

\begin{abstract}

We present a detailed computational study of the assembly of
protostellar disks and massive stars in molecular clouds with
supersonic turbulence.  We follow the evolution of large scale
filamentary structures in a cluster-forming clump down to protostellar
length scales by means of very highly resolved, 3D adaptive mesh
refined (AMR) simulations, and show how accretion disks and massive
stars form in such environments.  We find that an initially elongated
cloud core which has a slight spin from oblique shocks collapses first
to a filament and later develops a turbulent disk close to the center
of the filament.  The continued large scale flow that shocks with the
filament maintains the high density and pressure within it.  Material
within the cooling filament undergoes gravitational collapse and an
outside-in assembly of a massive protostar.  Our simulations show that
very high mass accretion rates of up to $10^{-2} \, \Msol \, \yr^{-1}$
and high, supersonic, infall velocities result from such filamentary
accretion.  Accretion at these rates is higher by an order of
magnitude than those found in semi-analytic studies, and can quench
the radiation field of a growing massive young star.  Our simulations
include a comprehensive set of the important chemical and radiative
processes such as cooling by molecular line emission, gas-dust
interaction, and radiative diffusion in the optical thick regime, as
well as $\Htwo$ formation and dissociation. Therefore, we are able to
probe, for the first time, the relevant physical phenomena on all
scales from those characterizing the clump down to protostellar core.

\end{abstract}



\section{Introduction}
\label{sec:intro}

Arguably the most comprehensive theoretical picture that we have of
star formation is that stars are a natural consequence of supersonic
turbulence within self-gravitating, molecular clouds \citep[see recent
reviews][]{Ballesteros06, Elmegreen04, MacLow04}.  It has been
appreciated for some time that turbulence could play an important role
in controlling the formation of cosmic structure in general and stars
in particular \citep[e.g.,][]{Weizsaecker51}.  Supersonic turbulence
is observed in most if not all giant molecular clouds (GMCs) and is
important because it rapidly sweeps up large volumes of gas and
compresses it into systems of dense filaments.  The density jump that
is associated with isothermal shock waves of Mach number $\Ma$ scales
as $\simeq \Ma^2$, so that this process may ultimately be responsible
for forming the dense clumps and cores seen in molecular clouds
\citep[e.g.,][]{Padoan95}.  An attractive aspect of this theory is
that since turbulence is associated with a broad spectrum of
velocities, a large range of of physical scales and masses should
arise for both the filamentary structure, as well as the properties of
the dense clumps and cores that form within them.

Observational surveys confirm that filamentary substructure
characterizes the internal organization of molecular clouds on many
scales.  It is clearly evident in the Orion A molecular cloud
\citep[e.g.,][]{Johnstone99} where, in addition to showing the obvious
integral-filament shaped structure, one also sees smaller structures
of $1.3 \, \pc$ in scale.  JCMT submillimeter studies of Orion B
\citep{Mitchell01} reveal a plethora of filamentary structures and
their embedded cores.  This pattern is also seen in the observational
studies by \citet{Fiege04}, wherein a few bright cores are seen to be
embedded in a larger, isolated filamentary structure.  Similar results
are also seen in more embedded regions such as the Lupus 3 cloud where
strong links between filaments and emerging star clusters are observed
\citep{Teixeira05}

Computational studies of turbulence \citep[e.g.,][]{Porter94} have
made important progress over the last decade both in their dynamic
range and in the list of physical processes that have been added to
them.  Simulations show that the shocks that are created in supersonic
turbulence produce a rich filamentary substructure.  A broad spectrum
of smaller, condensed regions is also produced
\citep[e.g.,][]{Klessen00, Padoan02}.  Detailed comparisons with the
observations show that there is a good correspondence between
numerically determined core mass spectra and the submillimeter surveys
of star forming cores in molecular clouds \citep[e.g.][]{Tilley04}
(henceforth TP04).  One of the most important consequences is that the
core mass spectrum (CMS) is found to be similar to that of the initial
mass function (IMF) that characterizes the stellar mass spectrum .
Cores in these numerical experiments are also born with a broad
distribution of spins, which are a consequence of the oblique shocks
that create them (e.g., TP04). The general velocity field
created in simulations of compressible gas has also been shown to have
a very good correspondence with that observed molecular cloud spectra
\citep{Falgarone94}.  Finally, structure formation in supersonic flows
occurs rapidly - typically in a sound crossing time or or two.  This
can in principal account for the rapid time-scales that characterize
the formation of star clusters such as the Orion Nebula Cluster (ONC)
- whose members have a mean age $ < 1 \Myr$ with a spread of less than
$2 \, \Myr$ \citep{Hillenbrand97}.  Thus, the universal properties of
turbulence potentially provides a comprehensive explanation for the
broad range in stellar masses and spins that characterize star
formation wherever we have been able to measure it.

The turbulent fragmentation theory of star formation makes a much
broader range of testable predictions than the older theory that
featured the gravitational collapse of nearly spherical, isolated
cores, often characterized as singular isothermal spheres.  This
latter picture did not provide a clear physical explanation for the
origin of the IMF.  Moreover, the model predicted that accretion rates
that are related to the cube of the (isothermal) sound speed $c$, i.e.
$\dot M_a \simeq c^3/G$.  While adequate to explain the formation of
low mass stars in atypical clouds such as Taurus, this accretion rate
is orders of magnitude too small to account for the formation of high
mass stars in observed regions of clustered star formation such as the
Orion Nebula \citep[see][for recent review]{Beuther06}.  It is known
that accretion rates of order $10^{-3} \, \Msol \, \yr^{-1}$ can
overcome the effects of radiation pressure and result in massive star
formation \citep{Wolfire87}.  It has been suggested that the cores in
which massive stars form can achieve these high rates essentially due
to the high turbulent pressure that is found in such regions
\citep[e.g.][]{McLaughlin97, McKee03}.

While high pressure must play an important role in the formation of
massive stars, these latter theoretical studies did not address how it
is that low mass stars would form in the high pressure region that is
producing the massive stars - both populations are co-spatial.
Moreover, it has been shown in simulations that the turbulence
envisaged to drive a high mass accretion rate onto a single massive
star will instead fragment the gas into smaller pieces quite rapidly
\citep{Dobbs05}.  Turbulence creates structure, and it is this
structure, not pressure, that ultimately determines mass flow.

A more fundamental shortcoming of the isolated core picture is that it
fails to explain how stars can span more than 3 decades in mass.
While the process of ambipolar diffusion - often invoked in older
models as a regulator of gravitational collapse of isolated magnetized
cores \citep[e.g.,][]{Shu87} - certainly does mediate the eventual
collapse of sufficiently massive cores, it does not provide a
first-principals explanation for why such a wide range of stellar
masses should exist in the first place.  Similarly, Jeans mass
arguments to "predict" stellar masses are not really very predictive
when applied to media that are dominated by large density contrasts.
As low and high mass stars form in clusters and are essentially
co-spatial they are likely to acquire their masses by similar
mechanisms. In both cases, we expect that filamentary accretion will
play a role in the accumulation of the cores out of which they form.

In this paper, we explore the multi-scale nature of the formation of
filamentary large scale structure and how this controls the formation
of protostellar disks and stars in turbulent cluster-forming clumps
(these have typical sizes of $< 1 \, \pc$.).  We accomplish this by using an
Adaptive Mesh Refinement (AMR) code known as FLASH \citep{FLASH00}.
Our 3D hydrodynamical AMR simulations include many physical processes
such as cooling due to dust, molecules, molecular hydrogen
dissociation, as well as radiative diffusion out of optically thick
regions.  We use as initial conditions numerical data from our recent
simulations of cluster formation in molecular clumps (TP04)
that show that a complete mass spectrum of gravitationally bound cores
can be formed by supersonic turbulence within molecular clouds that
closely resemble the core mass function of clouds such as Orion.  

We follow the filament that forms the first massive star in our
simulation, and find that it has high density and 
dynamic pressure that is maintained
by continued inflow and shocking into the filaments.  The collapse of
this dense material along the filament and into the disk and star is
the key to understanding this problem.  Lower mass stars in this
picture also form in filamented structure, but which are less
compressed and on smaller scales.  Our simulations trace how the most
massive star forming region is assembled and how collapse and the
formation of protostellar disks can occur, by resolving the local
Jeans length down to scales approaching that of the protostar itself.

We find that filaments play a dominant role in controlling the
physics, accretion rate, and angular momentum of the much
smaller-scale accretion disk that forms within such collapsing
structures \citep[see also][]{Balsara01b}. Large scale filamentary
flows sustain accretion rates that are orders of magnitude greater
than both (i) the naive scalings that are derived from the virial
theorem applied to uniform, 3D media, or even (ii) the collapse of
isolated Bonner-Ebert spheres.  The large scale structuring of
molecular clouds into filaments therefore has profound effects on the
rate of formation of disks and stars.

The formation and subsequent fragmentation of large scale filaments
can also lead to the formation of (massive) binaries or multiple star
systems which was shown in a series of SPH
simulations~\citep[e.g.][]{Turner95, Whitworth95, Bhattal98}. In those
studies, fragmentation is mainly the result of the complete formation
of molecular hydrogen (which reduces the temperature in the cloud
cores at high densities) during the gravitational collapse of the
large scale filament.
    
This paper proceeds by first describing our initial conditions and
numerical procedures (\S \ref{sec:methods}).  We then go on to
systematically study the evolution of large scale filaments (\S
\ref{sec:largescale}) that feed the formation of a massive star, the
formation and evolution of an accretion disk (\S \ref{sec:disk}), and 
finally the accretion of the central star (\S
\ref{sec:starformation}) within the disk. We conclude our results in
the discussion of (\S \ref{sec:discussion}) and give a compilation of
our cooling approach in the appendix~\ref{apx:cooling}.

\section{Numerical method and initial conditions}
\label{sec:methods}

Many studies of star formation have examined the self-similar collapse
of highly idealized structures such as singular isothermal spheres.
An initial state that is much closer to observational reality is the
3D a Bonner-Ebert sphere - whose collapse has received far less
theoretical attention until recently.  In \citet{Banerjee04} and
\citet{Banerjee06}, we adapted the FLASH code for star formation
applications and studied the three dimensional collapse of rotating,
magnetized and non-magnetized, Bonnor-Ebert-Spheres~\citep{Ebert55,
Bonnor56}.  We will compare our simulations with the results of the
collapse of isolate B-E models later in the paper.  As noted above
however, isolated objects like this are not typical structures of
observed molecular clouds.

\bef
\showone{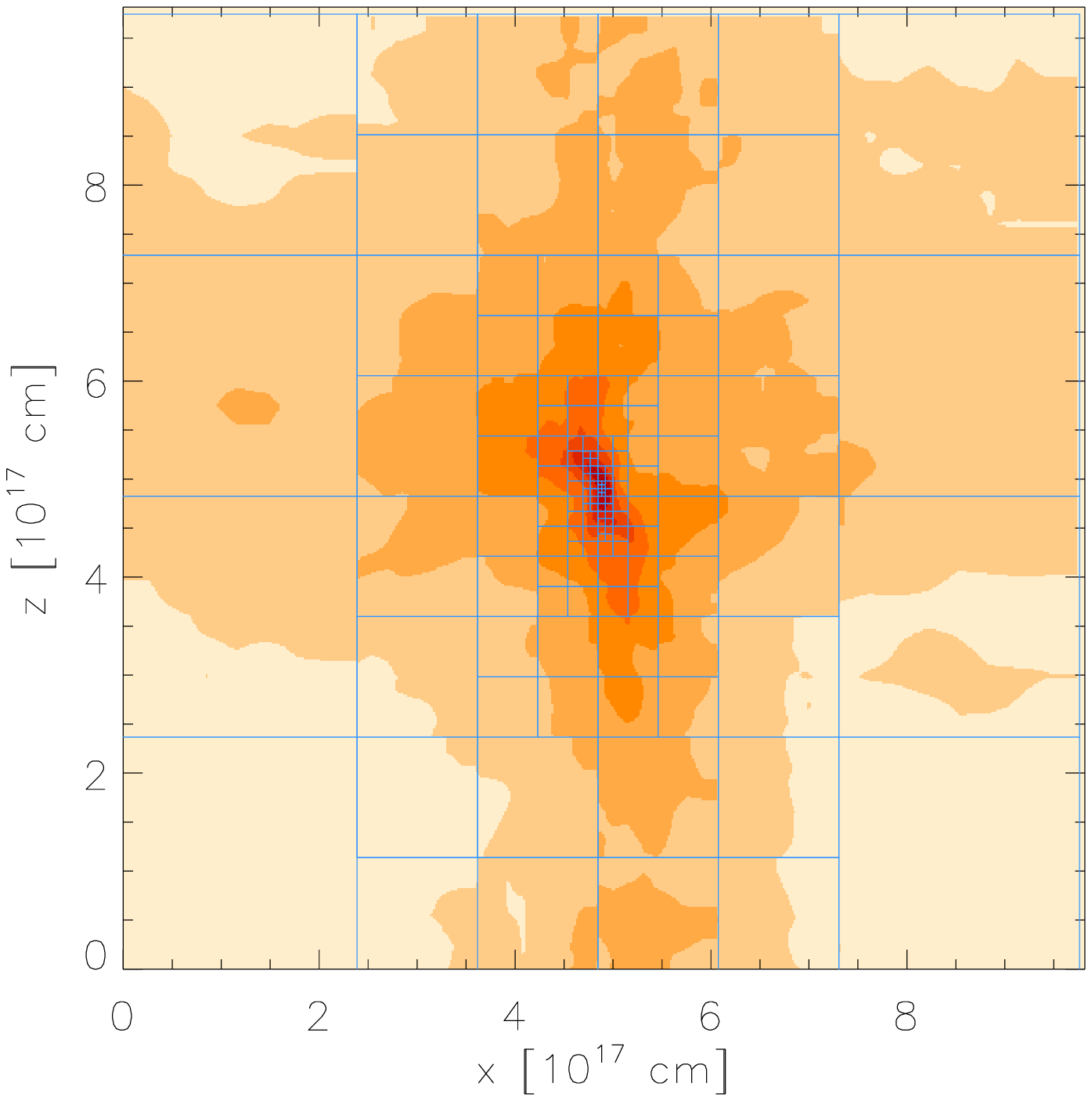}
\caption{Initial setup as ``seen'' by the FLASH code. The low density
  (density shown in gray scale)
  areas are less refined due to the applied Jeans refinement
  criterion. The resolution is indicated by the square blocks where
  each 3D block has $8^3$ grid points.}
\label{fig:initial}
\eef

This paper attacks the broader problem of the physical connection
between large scale structure formation (in this instance, on the
scale of the cluster forming clump) and star formation.  We use as
initial conditions the final state of simulations of turbulent core
formation presented by TP04.  The goal of this latter study was to
study the origin of the core mass function in cluster forming clumps,
and to identify the physical conditions that best matched the
observations.  As documented below, we found that there was a
particular set of conditions that produced realistic properties of
molecular cloud cores and that also abounded in filamentary-like
structures.

The TP04 simulations were hydrodynamic and were performed
with the ZEUS 3D code~\citep{Stone92a, Stone92b} on a static grid and
had to be stopped at the point when the density of any of formed cloud
cores reached the critical density of the Truelove
criterion~\citep{Truelove97}. Using the FLASH code~\citep{FLASH00}
which is based on the PARAMESH adaptive mesh refinement (AMR)
technique~\citep{PARAMESH99} allows us to follow the collapse of the
TP04 cores over many orders of magnitude in density
and length scale without under-resolving the physical relevant scales.

In practice we took the raw simulation data (velocity and density)
from one of the later output files in TP04, which are
available in HDF format, and initialized the FLASH variables with
these ZEUS-data. During the initialization we use a Jeans refinement
criterion where we resolve the local Jeans length,
\be
  \lambda_J = \left(\frac{\pi\,c^2}{G\,\rho}\right)^{1/2} \, ,
\label{eq:jeans_length}
\ee 
by at least 8 grid points ($c$ is the isothermal sound speed, $\rho$
is the matter density, and $G$ is Newton's constant).  This density
dependent initialization results in an initially non-homogeneous grid
structure compared to the homogeneous ZEUS 3D grid. By de-resolving
the low density regions in the simulations box we can save a
substantial amount of computational time without under-resolving the
physical scales which are governed by the Jeans length during the
collapsing stage. In Fig.~\ref{fig:initial} we show a 2D slice of
the initial grid structure as ``seen'' by the FLASH code.

For the subsequent simulation we increased the resolution of the Jeans
length to at least 12 grid points. Note, that the Truelove criterion
recommends that the Jeans length is resolved by only 4 grid points to
avoid numerical fragmentation and ensure convergence.

\subsection{Initial physical parameters}
\label{subsec:parameters}

For the initial setup, we take the late stage numerical data of run B5
from the TP04 simulations. This run has the following initial values
for the side length of the simulation box, $L = 9.8 \times 10^{17} \,
\cm = 0.32 \, \pc$, total mass, $M_{\sm{tot}} = 105.1 \, \Msol$,
isothermal sound speed, $c = 0.408 \, \km \, \sec^{-1}$, and a rms
velocity, $v_{\sm{rms}} = 5 \, c$ with a Kolmogorov-type spectrum,
respectively. The simulation run B5 resulted in a mass spectrum of
bound cores that has a Salpeter slope at high masses (in agreement
with the observations).  At the very highest masses, one finds a few
objects, the most massive of which was the first to collapse.  This is
the object whose collapse we continued to follow with the FLASH code
\footnote{For better visualization we shifted the all simulation
  variables so that the peak density appears at the simulation center
  before we re-run the FLASH simulation. This could be done without
  changing any physical properties as both simulations use periodic
  boundary conditions.}.  
At the time we re-started the collapse simulation with the FLASH code
the peak mass density is $1.35\times 10^{-15} \, \g \, \cm^{-3}$ which
corresponds to an initial free fall time, $t_{\sm{ff}} = \sqrt{3\pi/32
\, G \, \rho} \approx 1800 \, \ys$. The mass enclosed within a radius
of $2\times 10^{17} \, \cm$ around the peak density, which is
approximately the size of the cloud core, is $\sim 23 \, \Msol$. This
is also the Jeans mass of the initial TP04 simulation and represents a
fairly massive cloud core mass and its evolution can be compared to
the collapse simulation presented in \citet{Banerjee04} and
\citet{Banerjee06} which are based on the physical properties of a
$170 \, \Msol$ cloud core and the $2.1 \, \Msol$ cloud core Barnard 68
observed by~\citet{Alves01}.

\subsection{Cooling}
\label{subsec:cooling}

We extended the physical and chemical parameter space used in our
earlier work to incorporate the relevant cooling processes important
for star formation in molecular clouds. These processes are summarized
in complete detail in the appendix~\ref{apx:cooling} to which we refer
the reader for further information.  We included cooling by molecular
line emission, cooling from gas-dust interactions, radiative diffusion
in the optical thick regime, and cooling by $\Htwo$ dissociation.

For the remainder of this subsection, we give the more general reader
an idea of how we proceeded (one may wish to go directly to the
results in section \ref{sec:largescale}).  We first draw a distinction
between the optically thin and optically thick regimes.  For molecular
line emission in the optically thin limit, we used the cooling rates
due to \citet{Neufeld95}, as employed in \citet{Banerjee04} and
\citet{Banerjee06}.  Cooling by gas-dust transfer in the optically
thin regime was treated as in \citet{Goldsmith01}.  We show in the
appendix that the dust temperature approaches the gas temperature at
densities exceeding $n> 10^{7-9}$ under typical conditions.  Even
under these conditions however, we find that the dust is still a very
efficient coolant.

In the  optically thick regime, the radiation from the dust
surfaces (Eq.~\refeq{eq:dust_radiation}) cannot escape instantaneously
and, by definition, the mean free path of the radiation field
($\sim 1/\kappa$) becomes smaller than the typical core size of the
dense core. For most collapse situations the core size is given by the
core's Jeans length, $\lambda_J = \sqrt{\pi \, c_s^2 / G_N \,
\rho_{\sm{core}}}$. Therefore, we estimate the optical depth as
\be
  \tau \approx \kappa \, \lambda_J \, ,
\label{eq:optical_depth}
\ee
where $\kappa$ is the dust opacity.  This approximation is in
accordance with the results from opacity limited fragmentation and
star formation by \citet{Low76} and \citet{Silk77} who use the
expression above in their studies.

Using the fact that the dust temperature gets close to the gas
temperature (see section above) we can use Eqs.\refeq{eq:dust_opacity}
and \refeq{eq:optical_depth} to estimate the critical density at which
the gas is not cooled efficiently, i.e. when $\tau = 1$
\be
n_{\sm{crit}} = 2.1\times 10^{10} \,
  \left(\frac{T}{20\,\K}\right)^{-5} \,
  \cm^{-3} \,.
\label{eq:crit_dens}
\ee

In \citet{Banerjee04} we showed that this critical density can be used
to estimate the distance from the center at which the first shock
fronts build up. By updating our previous results with the above
critical density we find
\be
r_{\sm{crit}} \sim 17 \, \AU \,
  \left(\frac{n_{\sm{crit}}}{2.1\times 10^{10} \cm^{-3}}\right)^{-1/2} \,.
\ee
Test simulations of spherical collapse and the simulation reported
here confirm the appearance of first shocks at distances close to the
above value.

We treated the propagation of radiation in the optically thick regime
in the radiative diffusion limit (see the appendix).  We also
incorporated dust properties such as the frequency independence of
cooling at temperatures above $100 \, \K$, as well as the melting of
dust grains at temperatures around $1500 \, \K$.

Finally, we included both the formation and dissociation of molecular
hydrogen.  The formation rate of $\Htwo$ molecules on dust surfaces
was computed using the results of \citet{Hollenbach79}. We used the
dissociation rates of molecular hydrogen as computed in
\citet{Shapiro87}.

\bef
\showone{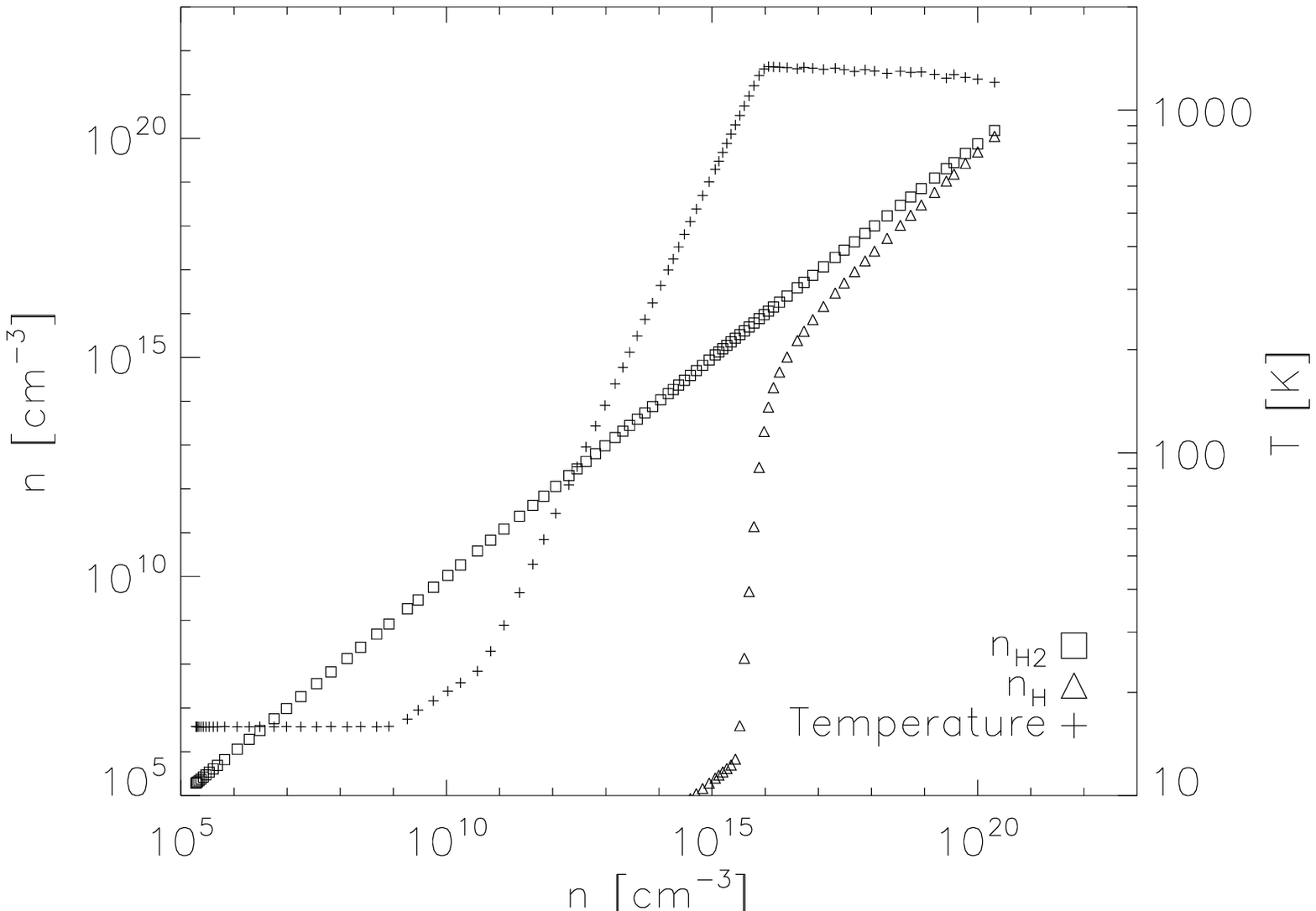}
\caption{Shows the evolution of the hydrogen densities and gas
  temperature of our non-rotating, non-magnetized collapse
  simulation. As predicted from our theoretical considerations the
  dissociation of $\Htwo$ molecules is a {\em self-regulating} process
  over a wide range of densities. The efficient cooling from the
  dissociation process (see Eq.\refeq{eq:h2_dissociation}) reduces the
  gas temperature and subsequently slowing down the dissociation of
  $\Htwo$ molecules.}
\label{fig:evol_simulation}
\eef

In order to understand the effects of these different processes in a
dynamic calculation, we first ran a series of test simulations of the
collapse of isolated, non-rotating, non-magnetized B-E spheres.
Fig.~\ref{fig:evol_simulation} shows the result of our BE-collapse
which include the process of $\Htwo$ dissociation (this simulation
started with the assumption that the gas is completely molecular to
begin with) .  As shown in the appendix, the temperature in the
radiation diffusion regime rises as $T \propto n^{1/3}$. At the
density $n_{\sm{cr}} \sim 10^{16} \, \cm^{-3}$, the temperature
reached $\sim 1200 \, \K$, which is hot enough to effectively
dissociate hydrogen molecules. However, the subsequent reduction of
the gas temperature reduces the dissociation efficiency leading to a
self-regulated, slow, disintegration of hydrogen molecules over a wide
range of densities.

We now turn to an analysis of the results of our turbulence simulations
and the formation of massive stars.

\section{Evolution of the large scale filaments}
\label{sec:largescale}

The initial state for our simulation, shown in Figure
\ref{fig:initial} (a 3D version of this is rendered in Figure 7 of
TP04), already shows that large scale structure has formed in our
simulation volume.  Hundreds of lower mass cores are present in this
volume, and as the virial analysis in that paper showed, many of the
cores have started to collapse.  The first core to violate the
Truelove criterion in TP04 is the most massive core.  As already
noted, the oblique shocks which are responsible for the core formation
are also responsible for the non-vanishing distribution of angular
momenta in the cores. As we will see later, this initial spin results
in a disk during the contraction phase.  The AMR simulation zeros in
on the evolution of the filamentary structure associated as it
collapses to form a disk and protostar.  The time-scale for the
development of this nonlinear, gravitationally dominated, structure is
very fast - amounting to several thousand years.  In this time frame
the larger filament scale remains fairly unaltered as material starts
to drain into the assembling protostellar disk.

\bef
\showone{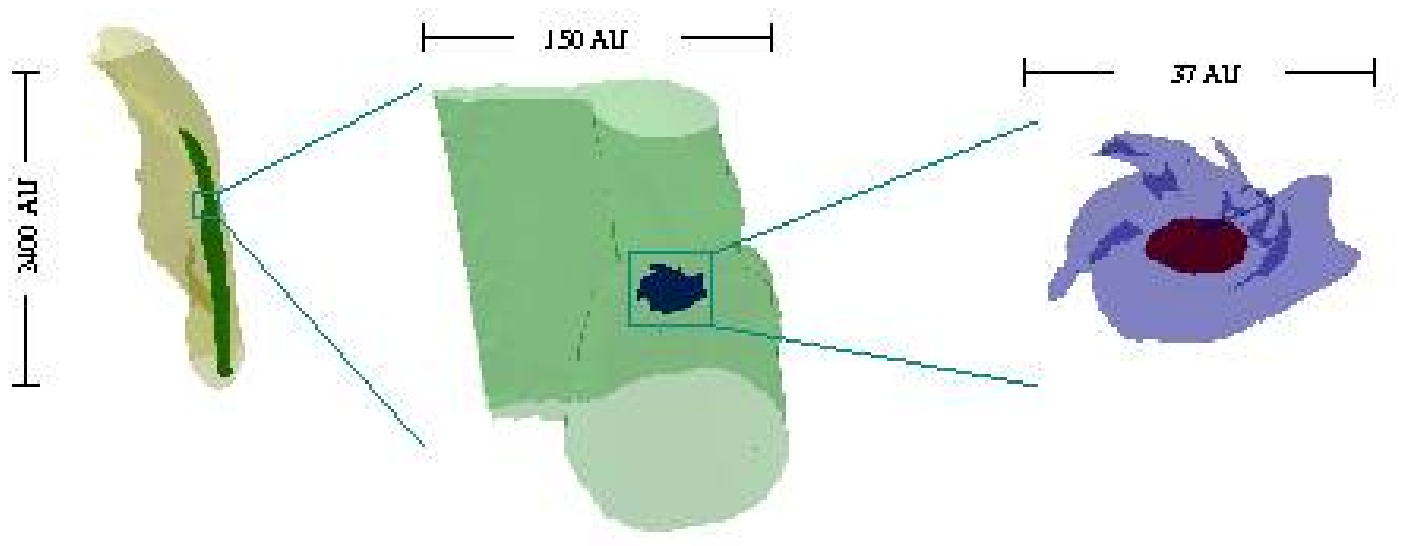}
\caption{Density isosurfaces at different scales: $2400 \, \AU$, $150
  \, \AU$, and $37 \, \AU$. The panels show the large scale filament,
  the protostellar disk within the filament, and the inner region of
  the disk. The density isosurface shown in opaque green in the left
  panel corresponds to the density shown in transparent green in the
  middle panel. Correspondingly, the opaque blue in the middle panel
  and the transparent blue in the right panel show the same density
  isosurface.}
\label{fig:3d}
\eef

The forming accretion disk is deeply embedded in the much larger scale
filament as is nicely shown in Figure~\ref{fig:3d}.  The large scale
filament is about $2600 \, \AU$ in length and one sees the converging
accretion flow on the $150 \, \AU$ scale view presented in the middle panel.  On
this intermediate, $150 \, \AU$ scale, the disk grows within a filament that
is developing out of a sheet-like structure.  As we will see, the
transfer of material from the sheet into the filament is associated
with net angular momentum - and tends to resemble the growth of a
large-scale, spinning vortex.  The disk, forming on yet smaller
scales, acquires this angular momentum from this larger scale process.
We see the actual disk at this time in the right panel, on a scale of
only $37 \, \AU$ in diameter.

\befone
\showone{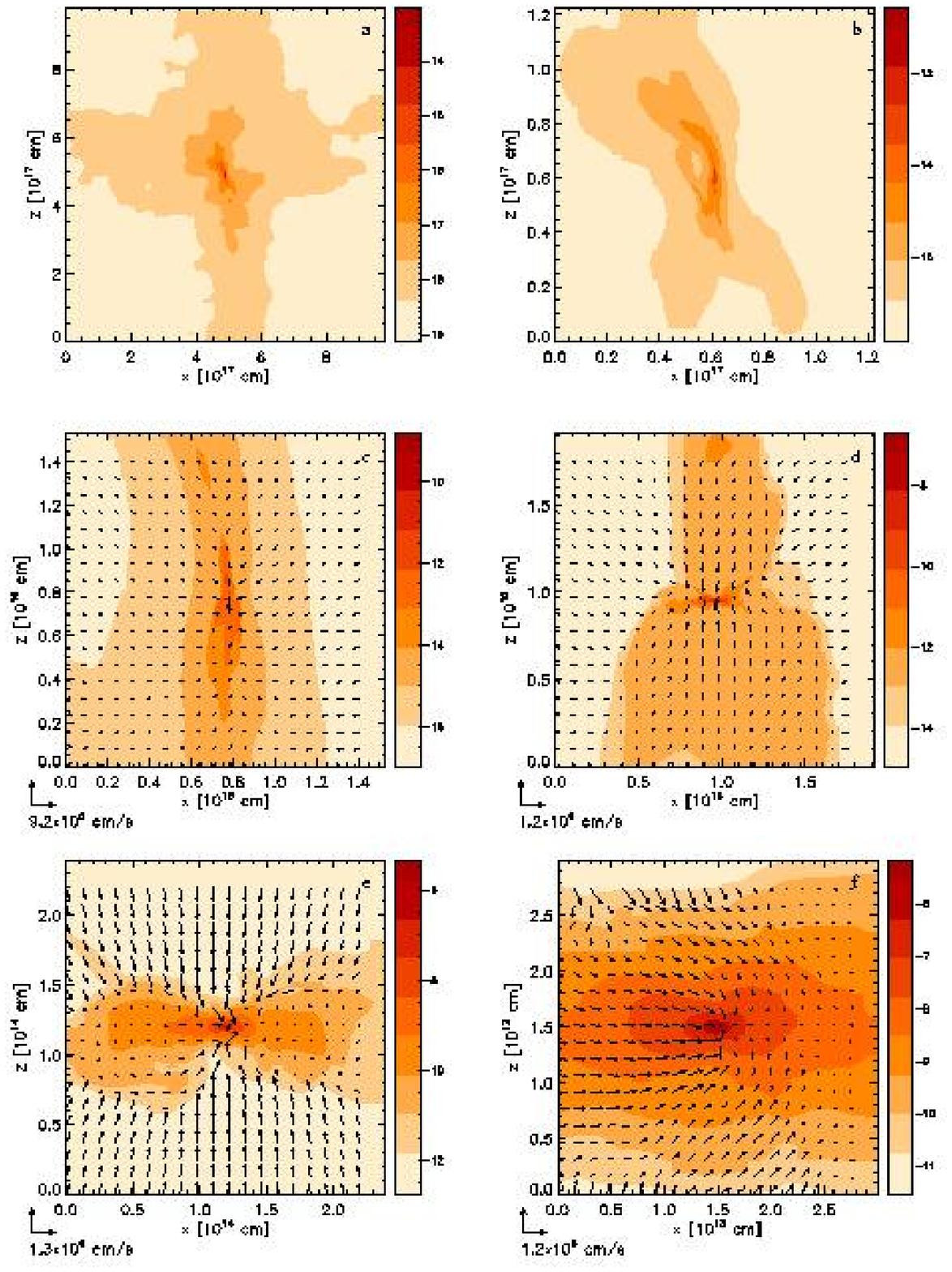}
\caption{Shows 2D slices of the matter density (logarithmic in grams
  per cubic centimeter) in the $xz$ plane at different scales from the
  latest stage of our simulation. The panels {\it a} to {\it f} show
  the density structure on continuously smaller scales: $0.3 \, \pc$
  (a), $8.2\times 10^3 \, \AU$ (b), $1\times 10^3 \, \AU$ (c), $128 \,
  \AU$ (d), $16 \, \AU$ (e), and $2 \, \AU$ (f). These slices cut
  parallel through the filament and show the protostellar disk edge
  on. The arrows indicate the velocity field.}
\label{fig:dens_slices_xz}
\eefone

\befone
\showone{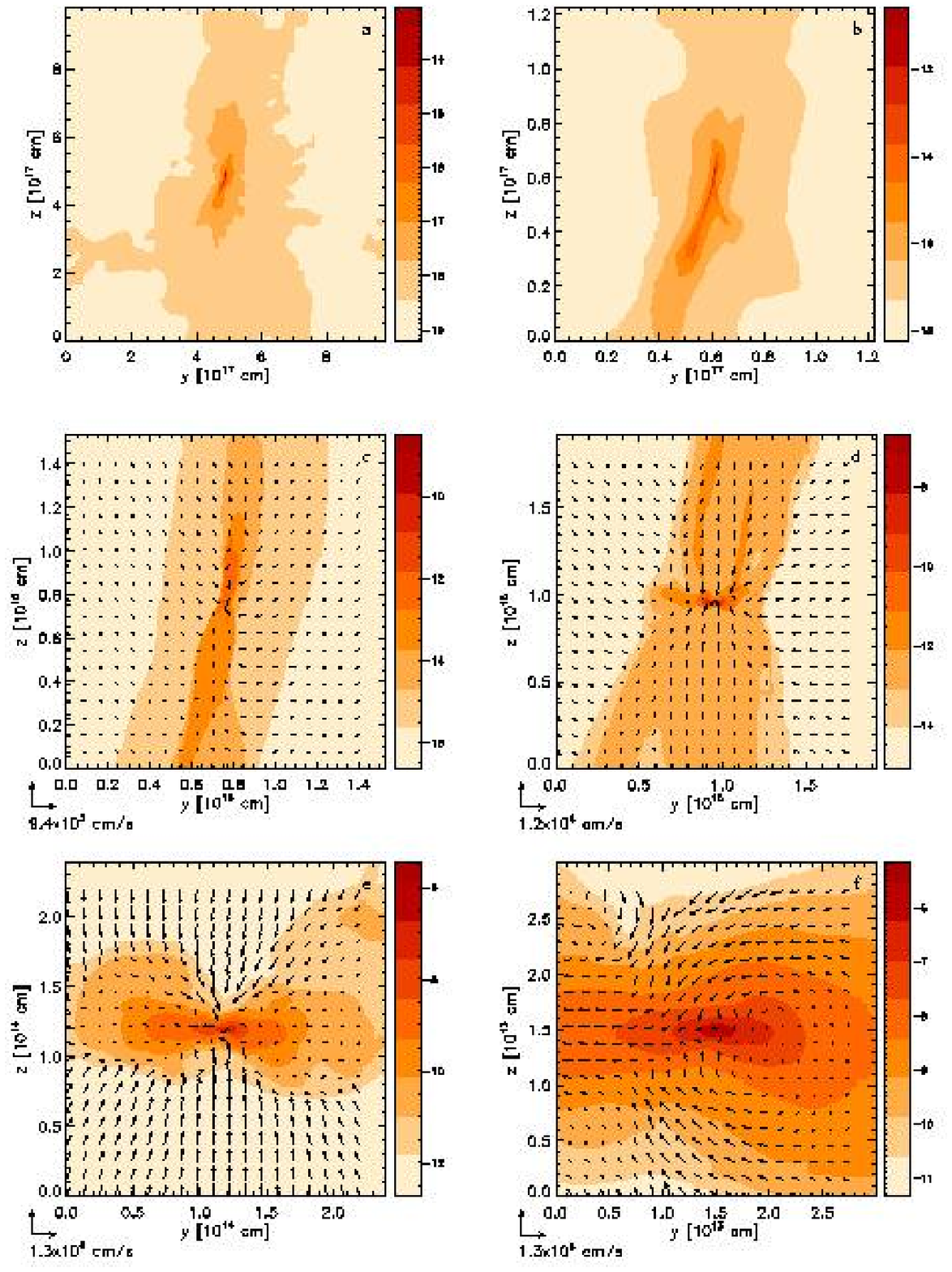}
\caption{Shows 2D slices of the matter density (logarithmic in grams
  per cubic centimeter) in the $yz$ plane at different scales from the
  latest stage of our simulation. The panels {\it a} to {\it f} show
  the density structure on continuously smaller scales: $0.3 \, \pc$
  (a), $8.2\times 10^3 \, \AU$ (b), $1\times 10^3 \, \AU$ (c), $128 \,
  \AU$ (d), $16 \, \AU$ (e), and $2 \, \AU$ (f). These slices cut
  parallel through the filament and show the protostellar disk edge
  on. The arrows indicate the velocity field.}
\label{fig:dens_slices_yz}
\eefone

\befone
\showone{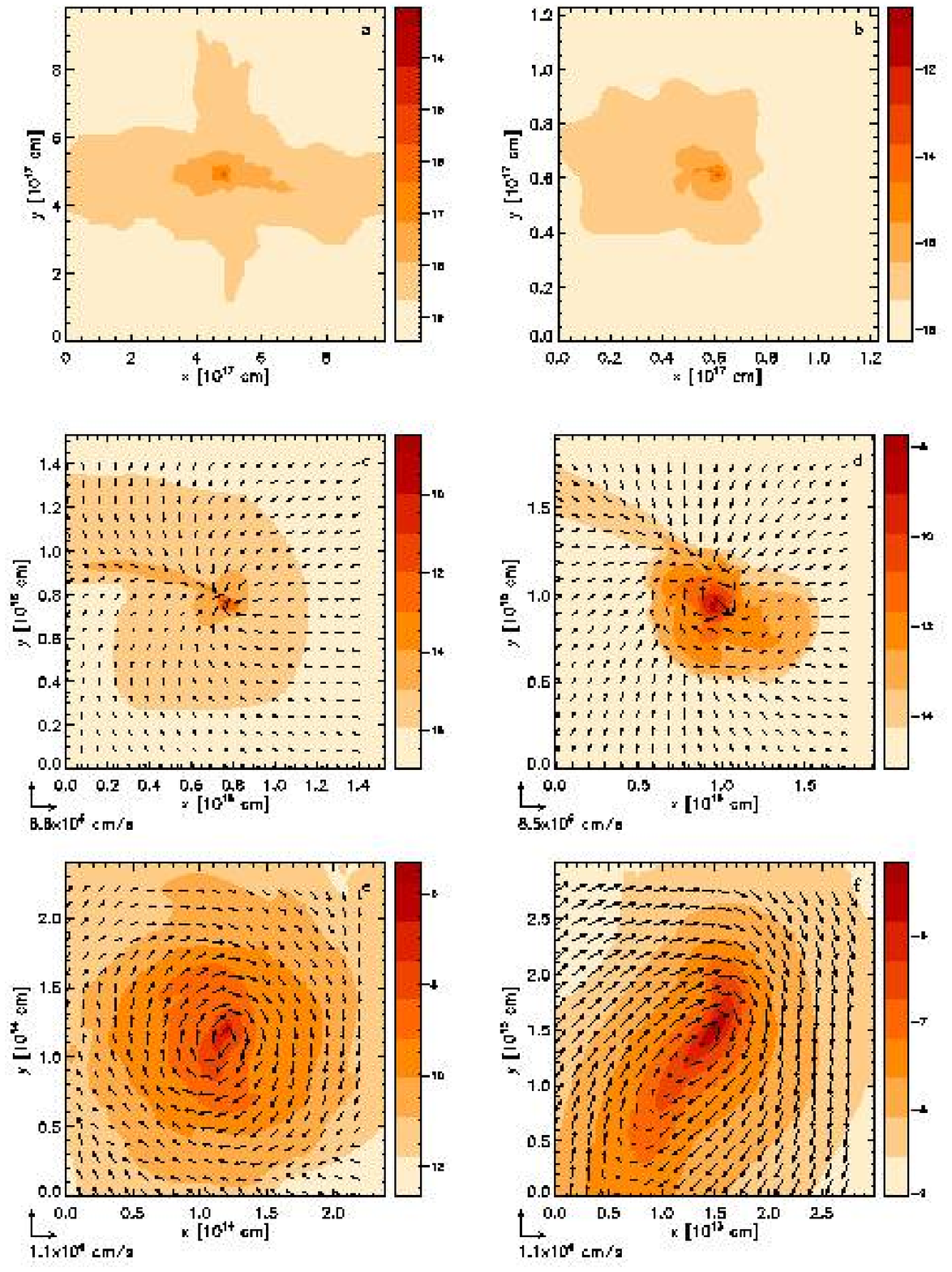}
\caption{Shows 2D slices of the matter density (logarithmic in grams
  per cubic centimeter) in the $xy$ plane at different scales from the
  latest stage of our simulation. The panels {\it a} to {\it f} show
  the density structure on continuously smaller scales: $0.3 \, \pc$
  (a), $8.2\times 10^3 \, \AU$ (b), $1\times 10^3 \, \AU$ (c), $128 \,
  \AU$ (d), $16 \, \AU$ (e), and $2 \, \AU$ (f). These slices cut
  perpendicular through the filament and parallel through the disk
  plane. The thin sheet attached to the filament is also clearly seen
  in the panels {\it c} and {\it d}. The arrows indicate the velocity
  field.}
\label{fig:dens_slices_xy}
\eefone

More details about the multi-scale structure of this filament are
shown in the three Figures~\ref{fig:dens_slices_xz} --
\ref{fig:dens_slices_xy}.  These represent 3 different 2D cuts through
the simulation data, at the end of our simulations.  The structures
are shown at increasingly resolved (finer) physical scales.
Figure~\ref{fig:dens_slices_xz} and~\ref{fig:dens_slices_yz} feature
cuts down the length of the filament which extend roughly along the
vertical ($z$) axis ( $xz$ and $yz$ planes respectively).  We clearly see a
resolved filamentary structure on the largest scales ($10^{17} -
10^{16} \, \cm$) scale in the upper left frame of the picture.  On the
smallest scales (highest resolution), shown in the bottom right panel,
we see the formation of a disk.

Two kinds of  large scale velocity fields are shown
in these figures: 

\noindent
(i) flow of material into the filament from larger scales.  This is a
consequence of the large scale velocity field associated with the
original supersonic turbulence field (most of the power is on the
largest scales - as in any general turbulent flow);

\noindent
(ii) large scale flow of material along the filament that converges to 
form a disk near to its mass center.    

In the first case, the large scale flow continues to shock the
filament and maintains the high pressure within it preventing its
dispersal.  We have measured the pressure in the initial state of this
filament to be of the order of $P/k_B \simeq 10^9 \, \K \, \cm^{-3}$,
which is high but expected in a converging flow at Mach numbers of
order 5.  In the second case, the material in the filament undergoes
gravitational collapse along the filament and towards the point of
concentration - the disk.
 
We show the disk (edge-on) close-up in the bottom panels of
Figs.~\ref{fig:dens_slices_xz} and \ref{fig:dens_slices_yz}. The
filamentary collapse flow undergoes a shock with the dense and slowly
cooling gas, and material joins the forming disk.  Given that this is
a highly asymmetric, 3D simulation, it is interesting that this small
scale disk is not badly warped, although it cannot be described as
being perfectly axisymmetric.

A top-down view of the forming disk, and another aspect of the
accretion flow that is giving its life, is seen in
Figure~\ref{fig:dens_slices_xy}. This cross-cut ($xy$ plane) through
the mid-plane of the disk shows that material from the large scale
filament and the attached sheet is accreting onto the disk. Panel {\it
d} of this figure especially shows an accreting structure that looks
like a picture of accretion flow in a Roche-overflow in a binary
system.  The disk at the highest resolution (panel {\it f}) shows the
presence of a bar, which is instrumental in driving a rather high
accretion rate through the disk, as we shall see.

While we do not follow the collapse of smaller regions in
Figure~\ref{fig:initial} in this paper, we know that filamentary
structure is associated with low mass cores as well (TP04).  These
varied filaments are the result of shocks on different physical scales
and different intensities.  This implies that the mass flow rates
through them and onto their embedded cores would be very
different. Thus, low mass stars would also be expected to form through the
filamentary accretion we study here.

\section{Disk formation}
\label{sec:disk}

We present in Figures~\ref{fig:temp_slices_z} and
\ref{fig:temp_slices_xy} a close-up snapshot of the temperature structure
of the disk, on $\sim 16 \, \AU$ scales.
Figure~\ref{fig:temp_slices_z} shows the accretion shock very clearly.
Material, raining into the disk from large scale flows along the
filament, shocks with the disk because the gas cannot cool quickly
enough at these densities.  One notes that rather high temperatures of
the order of $1000 \, \K$ are associated with this accretion shock
which envelops the entire disk. The post-shock gas cools rather
quickly, and is found at much more modest temperatures of $200 - 400
\, \K$. As demonstrated in \citet{Banerjee04} and by others
\citep[e.g.,][]{Yorke95,Tomisaka02,Matsumoto04} this cooler
post-shocked gas gets re-accelerated and shocks again further into the
gravitational potential resulting in a double (or multiple) shock
structure of the accreting gas.

A top-down close-up of the disk's temperature distribution is shown in
Figure~\ref{fig:temp_slices_xy}. We clearly see a well resolved large
scale spiral wave pattern with two spiral arms (on $16 \, \AU$
scales), that appears to be attached to a bar at the center of the
disk.  This is the state of the disk at the earliest times, long
before the Class 0 phase has ended.  The presence of this density wave
is central to the rapid transport of disk angular momentum that we
shall examine later.

\befone
\showtwo{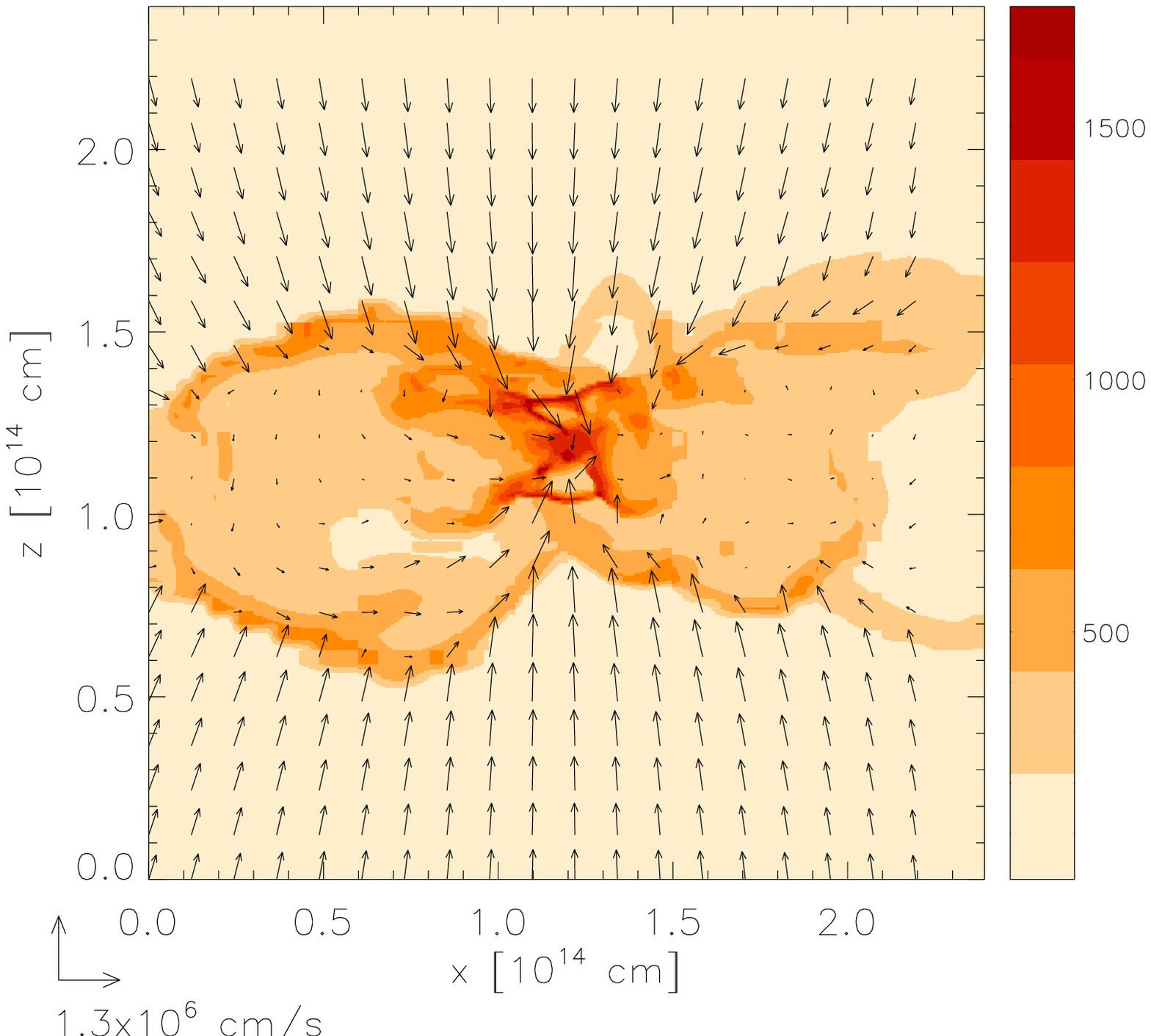}
	{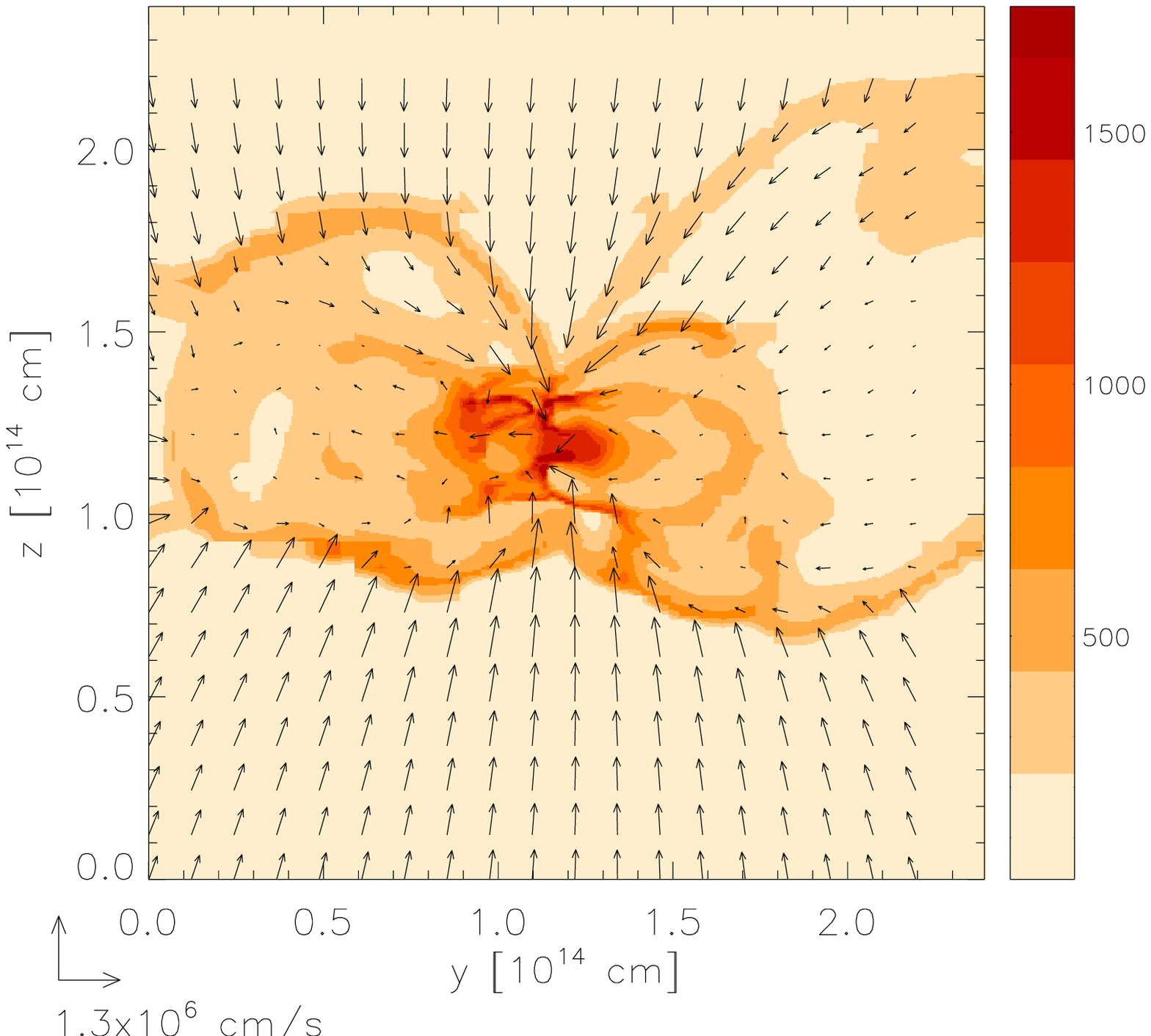}
\caption{Temperature contours (in Kelvin) in the $xz$ and $yz$ plane
  which cut perpendicular to the disk plane (scale: $\sim 16 \,
  \AU$). The first shocks appear in the region of the optical thin --
  optical thick transition. The arrows indicate the velocity
  field. The corresponding density maps are shown in panel {\it e} of
  Fig.~\ref{fig:dens_slices_xz} and Fig.~\ref{fig:dens_slices_yz},
  respectively.}
\label{fig:temp_slices_z}
\eefone

In order to understand what kind of emission one might expect from
such a disk (in accordance with our radiation diffusion
approximation), we present an optical depth map - which maps out
values of the Rosseland mean opacity of the disk - is shown in the two
panels of Figure~\ref{fig:tau_slices}.  We see the disk in
edge-on as well as a top-down views.  The post shock gas is very dense
and the grains give it a high optical depth ranging from $\tau \simeq
10^3 - 10^6$ in the dense regions on $\sim 10 \, \AU$ scales.  One
also sees that the spiral wave features seen in the temperature and
density maps also show up in the optical depth maps of the disk's
radial structure.

\bef
\showone{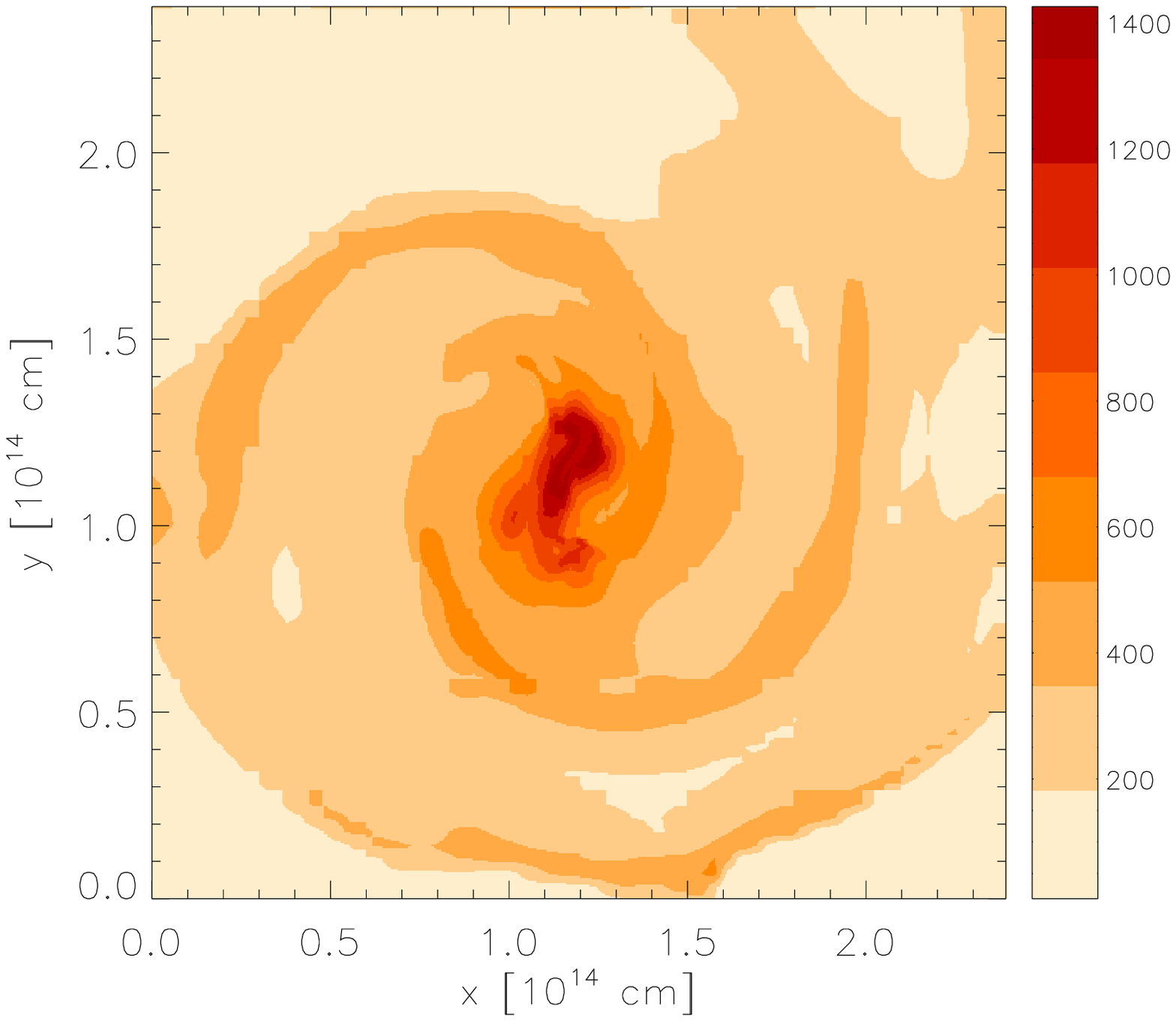}
\caption{Temperature contours (in Kelvin) in the disk plane
   ($xy$-plane) at $\sim 16 \, \AU$ scales. Clearly visible are the
   two arm spirals and the center bar in this temperature map. The
   corresponding density map is shown in panel {\it e} of
   Fig.~\ref{fig:dens_slices_xy}.}
\label{fig:temp_slices_xy}
\eef

\befone
\showtwo{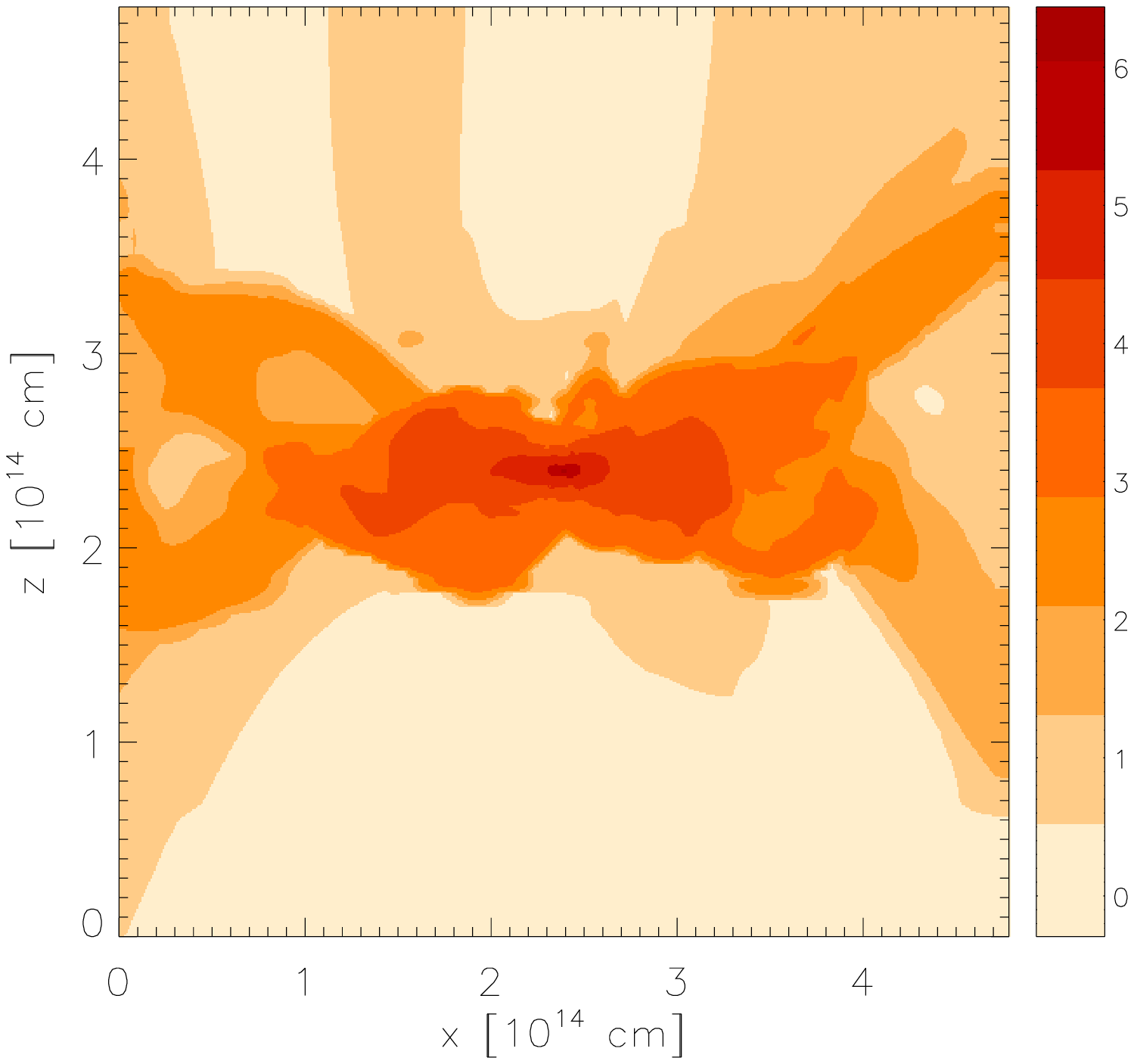}
        {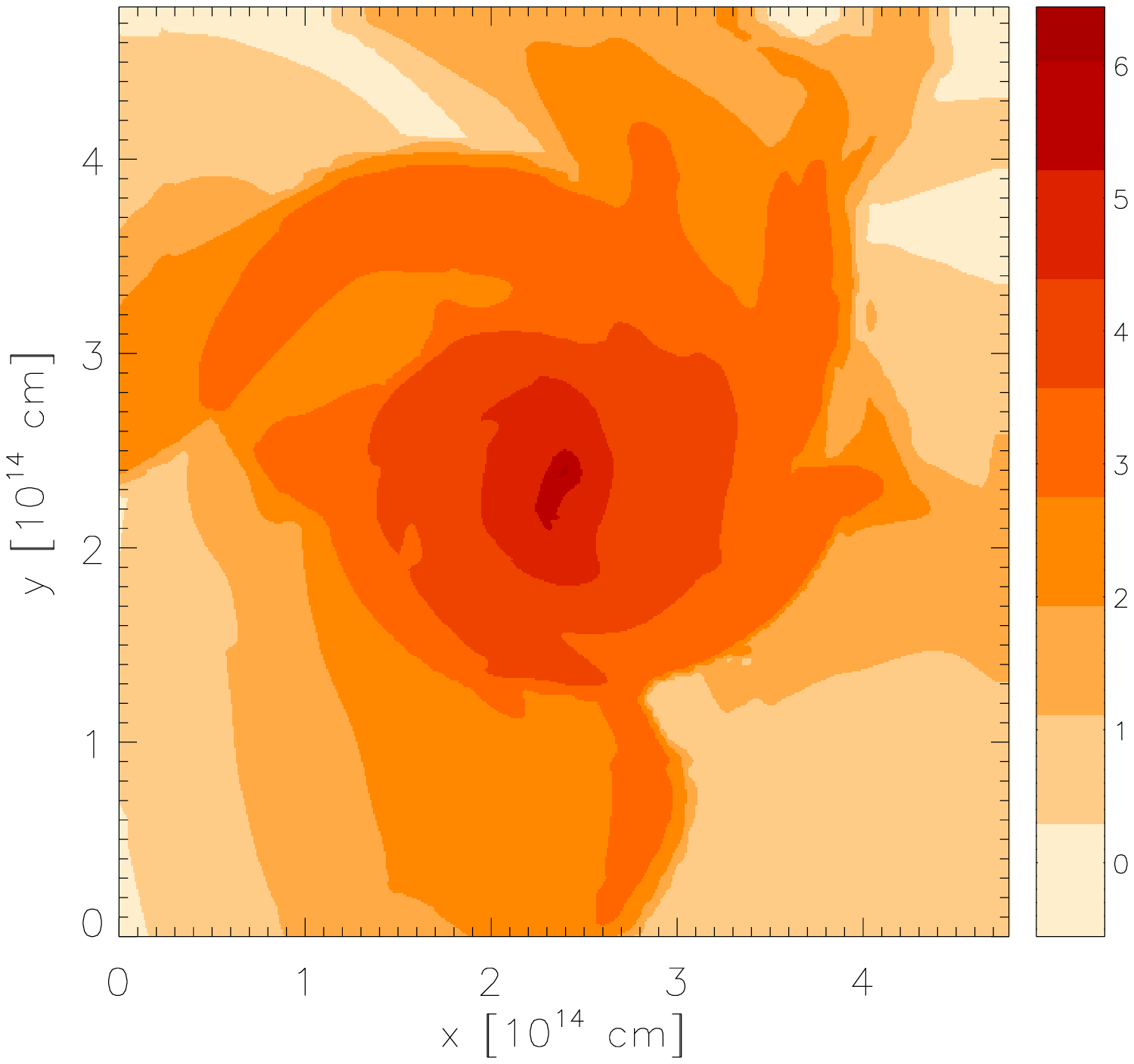}
\caption{Optical depth ($\tau$ logarithmic see
  Eq.\refeq{eq:optical_depth}) perpendicular to the disk plane (left
  panel) and in the disk plane (right panel) at $\sim 32 \, \AU$
  scales. The optical depth varies over six orders of magnitude within
  the protostellar disk and resembles its spiral structure.}
\label{fig:tau_slices}
\eefone

The evolution of the radial structure of the disk is shown in
Figures~\ref{fig:mass_evol} and \ref{fig:dens_evol}~\footnote{These
radial profiles show spherically averaged quantities with the center
at the peak density}. In these very early and deeply embedded 
stages of disk and star
formation, we find that the disk is actually more massive than the central
protostellar object for a brief period.  This is seen in the left panel of
Figure~\ref{fig:mass_evol} wherein the mass inside of $1 \, \AU$ at
the latest time is roughly $10^{-2} \Msol$, whereas the mass enclosed
within $10 \, \AU$ is about 10 times larger; $10^{-1} \Msol$.  Another
interesting feature of this mass distribution is that the mass within
the disk and protostar is a small fraction of the total mass that is
distributed on larger scales.  At $10^{17}$ cm, this amounts to a
total of $13 \, \Msol$ while at a parsec scale, we have a total of 100
solar masses - reflecting the mass of the cluster forming clump.  This
distribution of mass is very similar in character to simulations of
the first star formed in the universe, by~\citet{Abel02}.  Although the
cooling function of gas forming the first star is - of course - vastly
different than later epochs that we are simulating here, nevertheless
the two results both show that massive stars are like the tips of
icebergs in much more massive and extensive mass distributions.

The right hand panel of Figure~\ref{fig:mass_evol} shows that
the central density of the disk  grows very rapidly with time.  At
each time, the run of density has a central, flat core-radius (which
size is of the order of the core's Jeans length) with a steeply
falling density dependence beyond.  Variations in the power-law slope
arise because of the onset of different cooling rates of
material at different densities.  

The dominance of the disk over the protostellar mass at this early
time does not violate any observations of disk-star systems since the
latter are generally available typically for Class I TTSs and later.  Theoretical
calculations of disk formation often employ the assumption of
self-similar collapse, in which the central object dominates the mass.
We do not see this in our simulations.  We also note that a
dominant central mass does not appear in our earlier simulations of
the hydrodynamic collapse of a Bonner-Ebert sphere \citep{Banerjee04}.

\bef
\showtwo{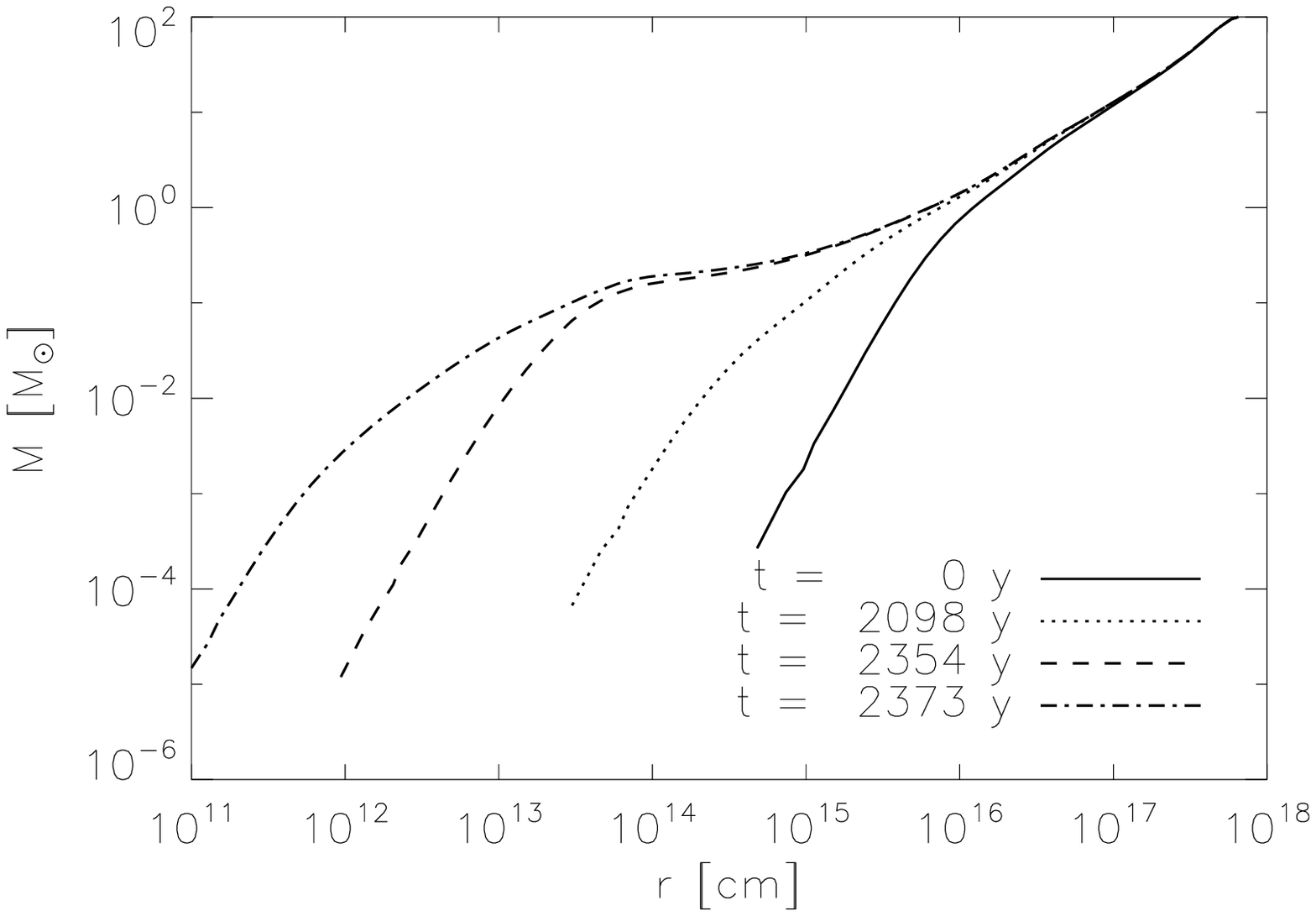}
        {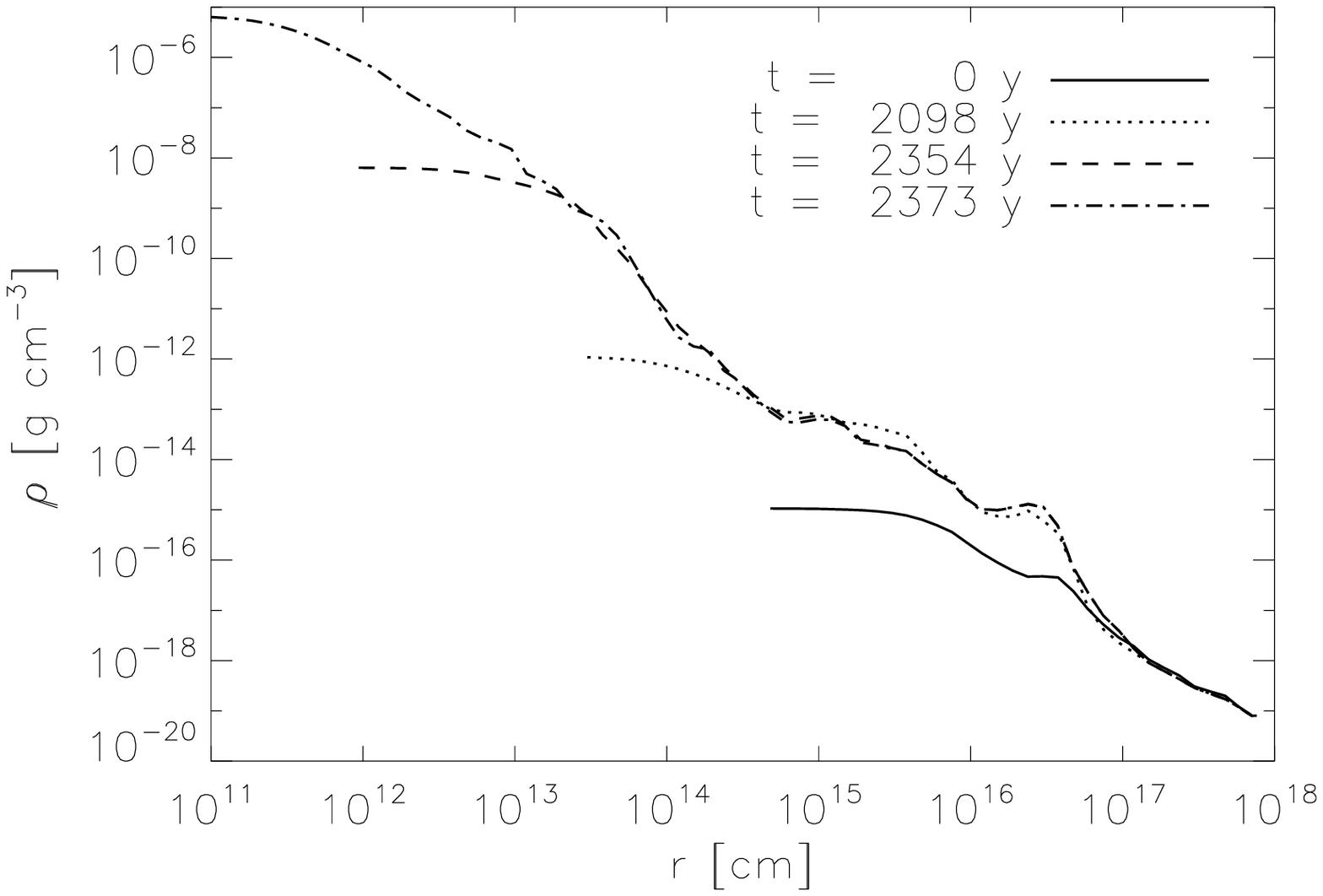}
\caption{Radial profiles of the enclosed mass and mass density at
  different times. The $t = 0 \, \ys$ graph shows the profile at the
  beginning of our simulation run and the subsequent labels indicate
  the time past since then.}
\label{fig:mass_evol}
\eef

\bef
\showtwo{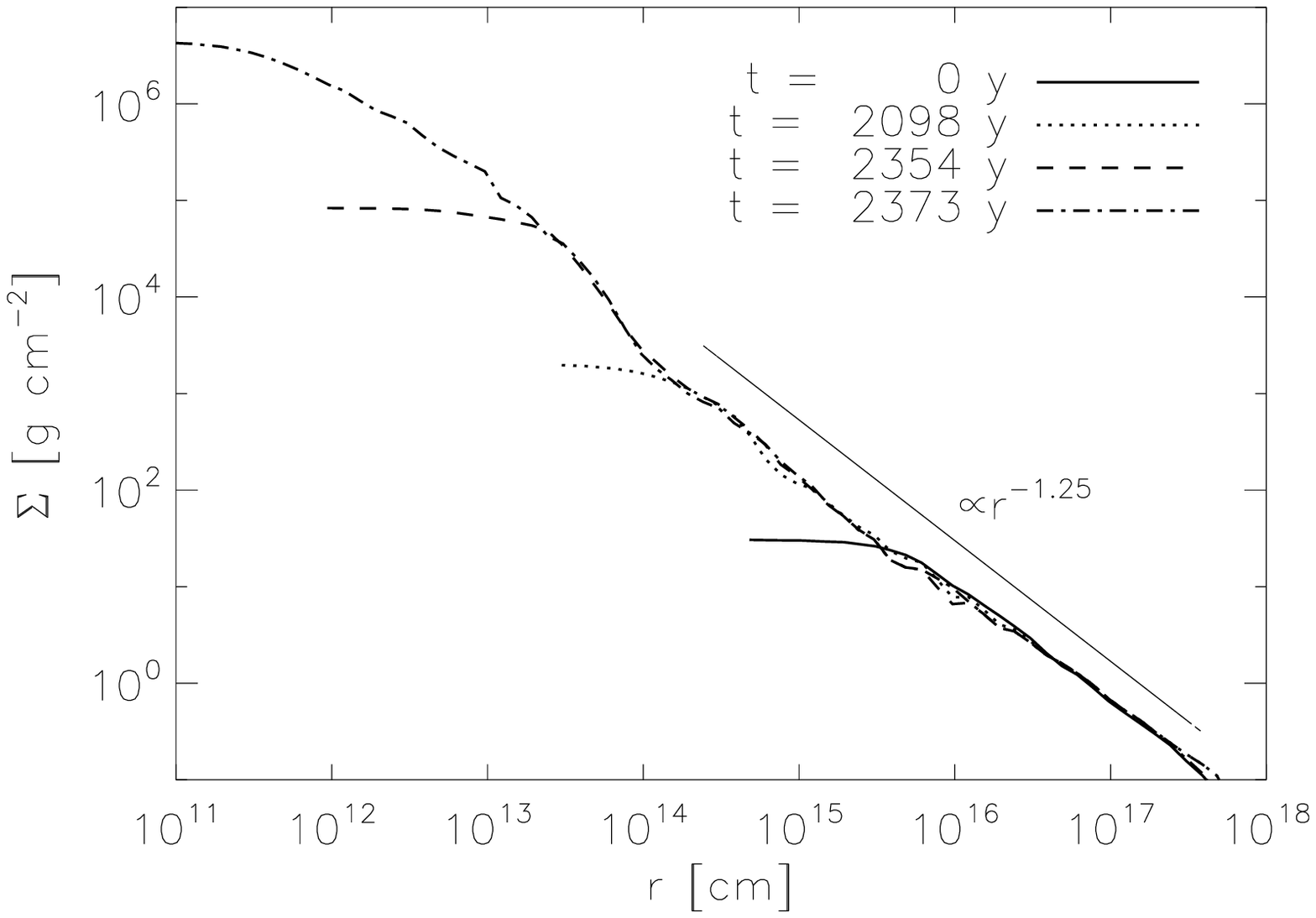}
        {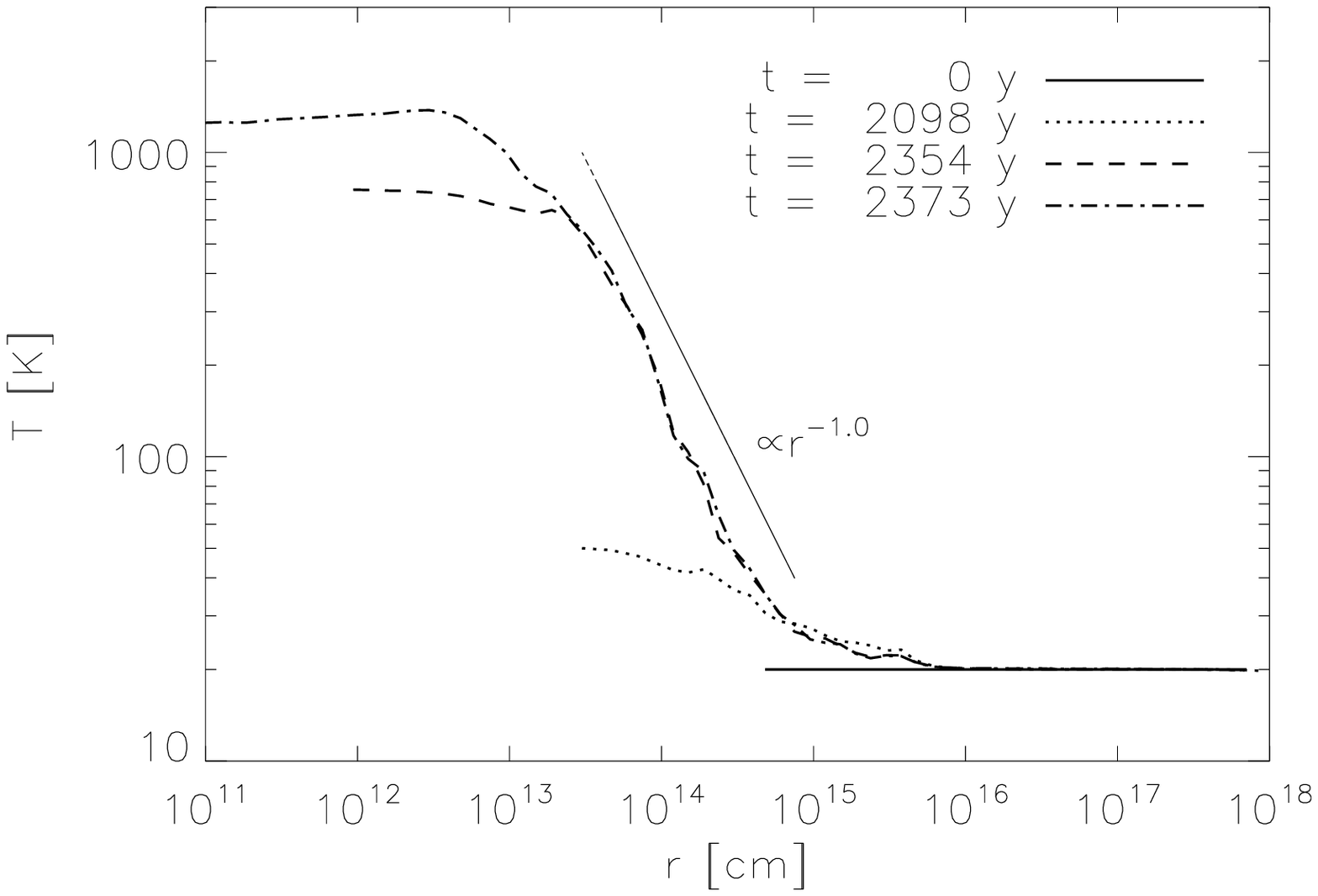}
\caption{Radial profiles of the surface density and temperature at
  different times. The $t = 0 \, \ys$ graph shows the profile at the
  beginning of our simulation run and the subsequent labels indicate
  the time past since then.}
\label{fig:dens_evol}
\eef

Theoretical discussions of disk formation also often noted that disks
needed to be lower in mass than their central stars, otherwise they
would be out of equilibrium.  This presents no problem to these early
systems.  In fact, rapid stellar formation would be highly favored if
disks were more massive in their earlier stages.  These highly
unstable disks favor the appearance of high formation rates -- which
is what we will show later in this paper. 

We also do not see further fragmentation of the protostellar disk at
this stage. The increasing pressure of this optical thick disk
stabilizes the disk against fragmentation. This shows again that the
thermal evolution of the collapsing cloud core plays a crucial role in
determining the fate of the protostellar system. For instance,
molecular clouds will fragment easily in the regime where abundant
molecular hydrogen is formed and subsequently the pressure decreases
\citep[e.g.,][]{Turner95, Whitworth95,Bhattal98}.

Figure~\ref{fig:dens_evol} summarizes the evolution of the disk's radial
column density and temperature distributions.  The outer regions of the disk
beyond $10 \, \AU$ or so show a rather shallow column density profile,
$\Sigma \propto r^{-1.25}$ compared to the minimum mass solar
nebula. We also see that a high column density of $\simeq 10^3 \, \g
\, \cm^{-3}$ is achieved at about an $\AU$ scale, which is then
surpassed.  We are charting the non-equilibrium early phase of disk
formation and evolution just before the major delivery of mass into
the central protostar has occurred. It is therefore not a surprise
that transitory column densities in excess of the solar nebula model
will arise.  The radial temperature distribution in the disk scales as $T
\propto r^{-1}$ at larger radii, with a flat distribution in the disk
center. This steep temperature profile reflects again the early stage
during the assembly of the protostar where shocks from the infalling
gas onto the adiabatic core are still present.
It is interesting that while the
envelop of this collapsing region features relatively smooth variations 
of the column and density profiles, the temperature profile falls
steeply and then abruptly changes to the nearly isothermal
background.  This transition in disk temperature 
behavior may be the best way of detecting the 
outer accretion shock and hence the outer disk edge.

The question of how angular momentum is distributed and transported in
these early phases is arguably one of the most interesting and
important issues for disk evolution, and massive star formation.  We
plot, in the left panel of Figure~\ref{fig:jabs_evol}, the specific
angular momentum $j(r)$ (angular momentum per unit mass) as a function
of disk radius for different times.  We see that for all times after
the initial state, that $j(r)$ is an increasing function of disk
radius.  Rayleigh's theorem informs us that this is in fact a stable
distribution of angular momentum (rotating bodies with constant or
increasing specific angular momenta are stable).  Only one graph is
different; this corresponds to the initial state where we see a
leveling of $j(r)$ at radii $10^{15}-10^{16}$ cm.  Since the MRI
instability is not present here, we must conclude that the bulk of the
disk must be relatively free from turbulence in this early stage.

In the right hand panel of Figure~\ref{fig:jabs_evol}, we replot this
data in another interesting way - now showing the specific angular
momentum as a function of the mass enclosed.  This is a good way of
measuring how much angular momentum is extracted and transported away
from the enclosed mass \citep[e.g.][]{Abel02}. The figure shows that
the specific angular momentum of material within the $0.01 \, \Msol$
mass shell has quickly lost nearly 3 orders of magnitude of specific
angular momentum, having fallen from an initial value of $10^{20} \,
\cm^2 \, \sec^{-1}$ to about $10^{18} \, \cm^2 \, \sec^{-1}$ in only
$2400 \, \yr$.  This angular momentum transport is achieved by
the spiral waves that we have seen in the earlier Figures
(Figs.~\ref{fig:dens_slices_xy}, \ref{fig:temp_slices_xy}, and
\ref{fig:tau_slices}). As is well known, spiral wave torques are very
efficient in transporting disk angular momentum because of their long
lever arm.  In the absence of either outflows or turbulence (through
the MRI instability that would be active in the presence of a magnetic
field), the plot shows that the non-equilibrium disk is very efficient
in generating a spiral wave structure that efficiently transports angular
momentum out of the central disk.

\bef
\showtwo{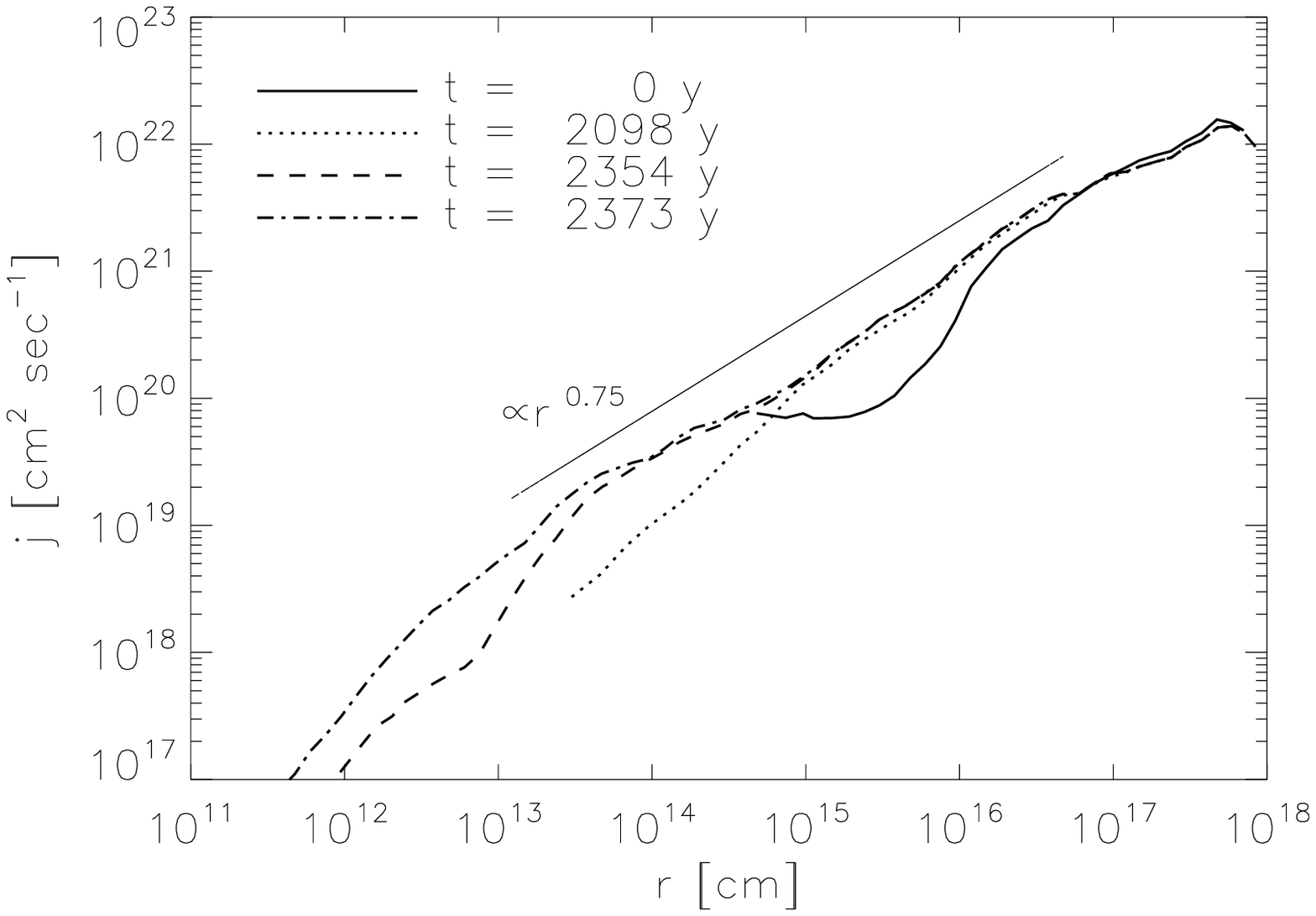}
        {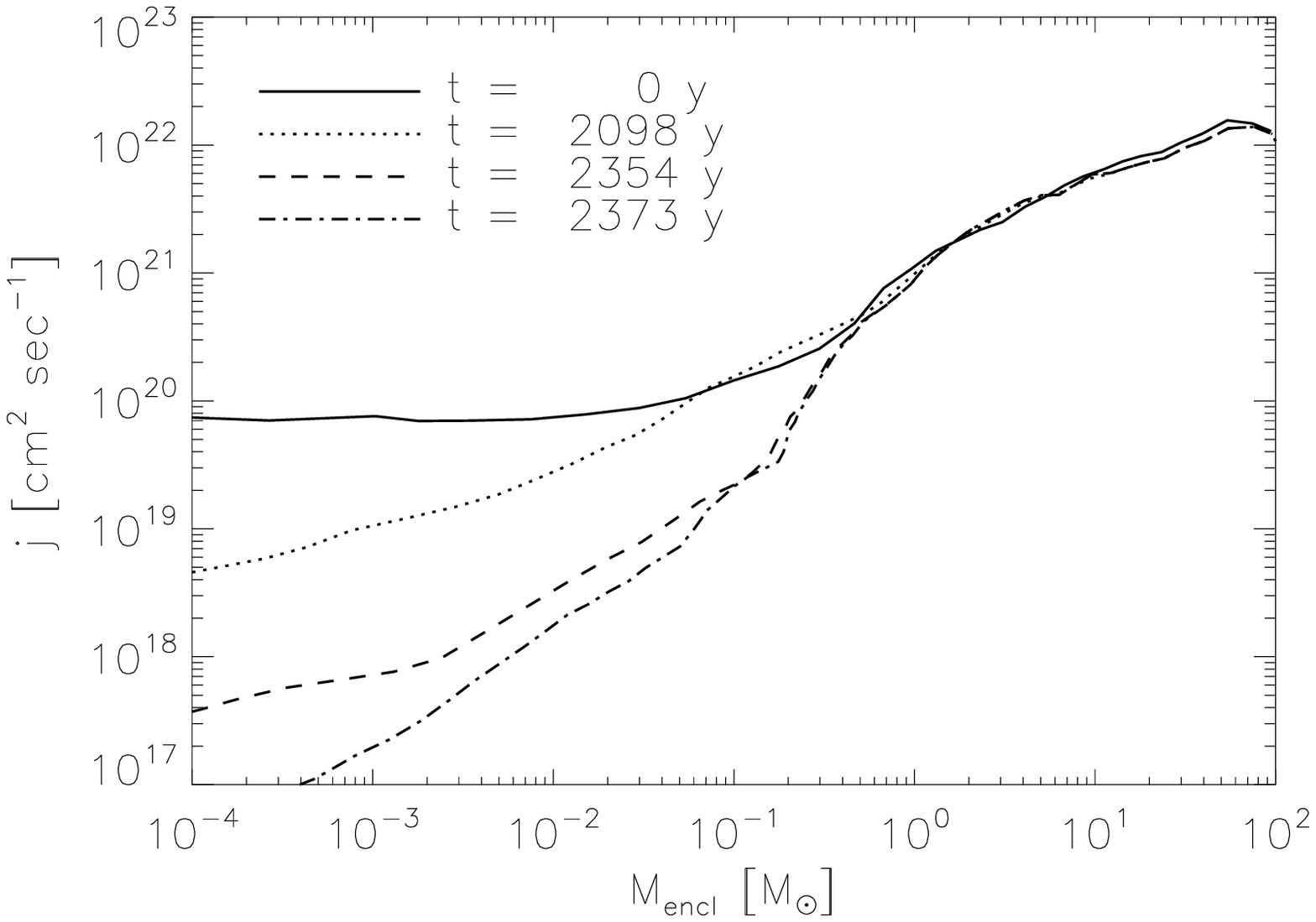}
\caption{The evolution of the spherical averaged specific angular
  momentum as a function of radius (left panel) and enclosed mass
  (right panel). During the accretion phase angular momenta is
  exchanged between ``mass shells''. The $t = 0 \, \ys$ graph shows
  the graphs at the beginning of our simulation run and the subsequent
  labels indicate the time past since then.}
\label{fig:jabs_evol}
\eef

\bef
\showtwo{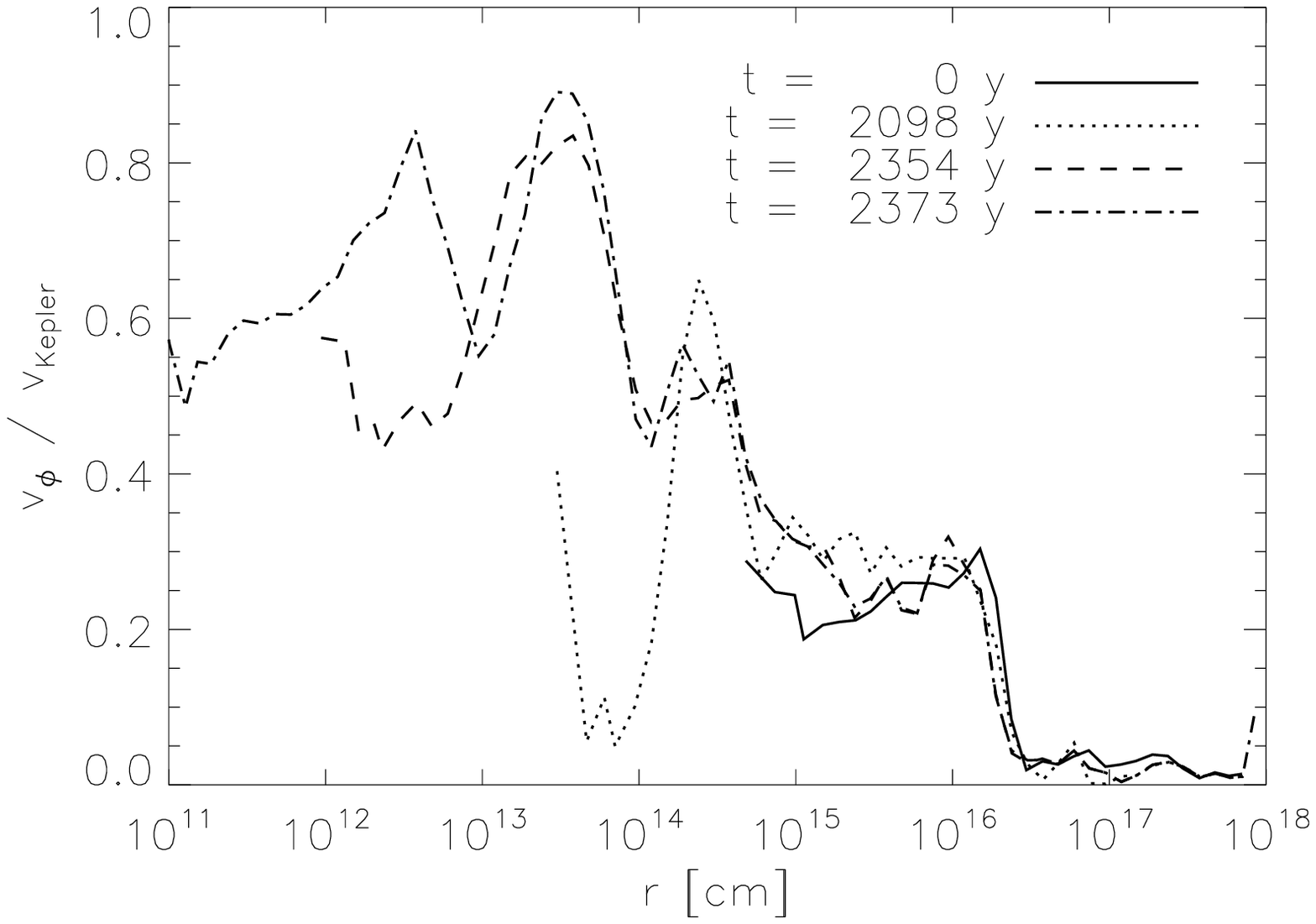}
        {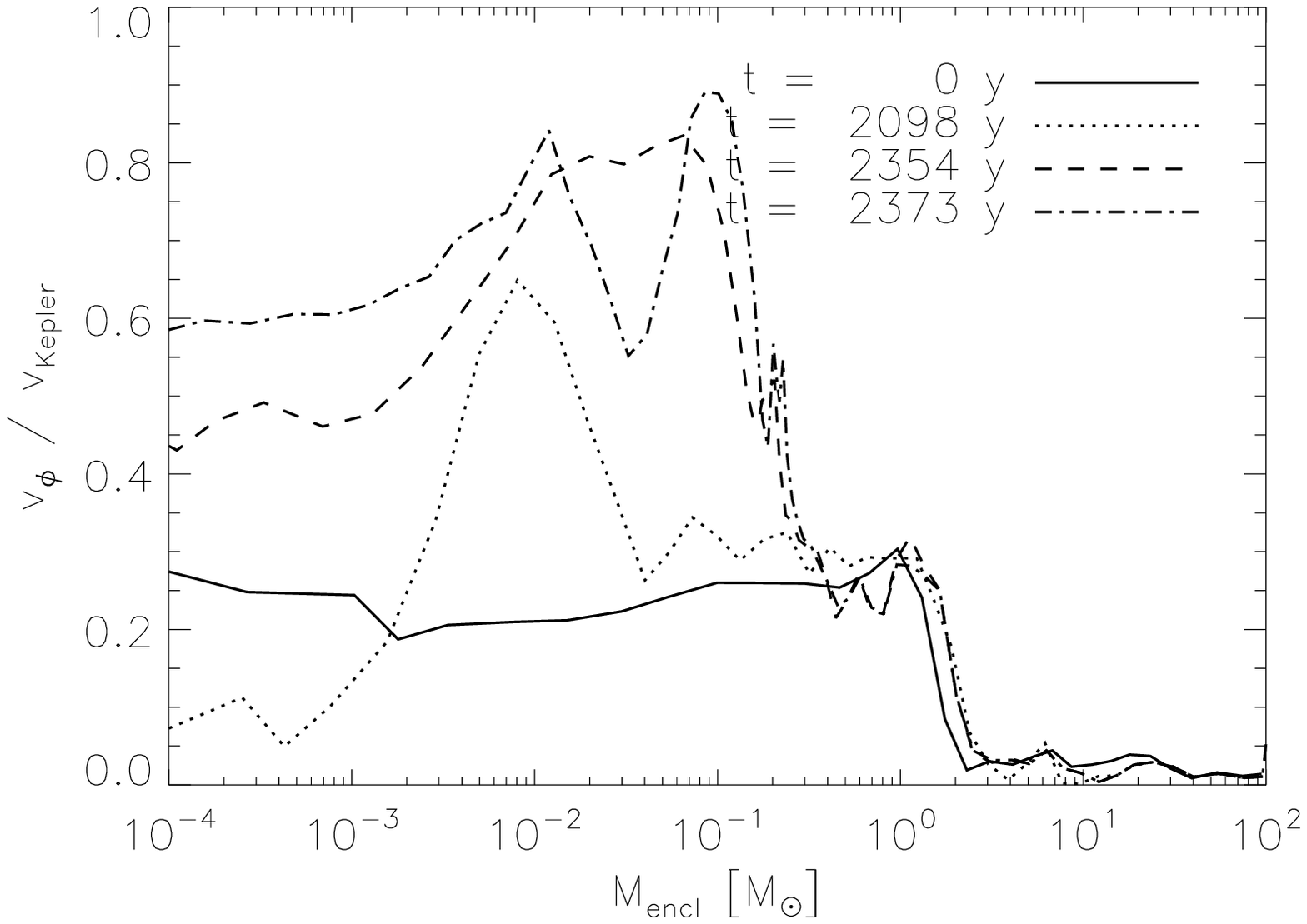}
\caption{The evolution of the rotational velocity compared to the
  Keplerian velocity as a function of radius (left panel) and enclosed
  mass (right panel). The $t = 0 \, \ys$ graph shows the profile at the
  beginning of our simulation run and the subsequent labels indicate
  the time past since then.}
\label{fig:vphi_evol}
\eef

We measure how far the disk is from a Keplerian distribution in
Figure~\ref{fig:vphi_evol}.  We normalize the measured rotation speed with
the local Kepler velocity by using the mass in the interior radius.
The left hand panel of this figure shows that this disk rotates more
slowly by a factor of about 2 than Keplerian value within several 10s
of $\AU$.  There are several peaks in the value of the rotation speed
suggesting that shock waves within the disk are active and may be
playing a role in helping to redistribute disk angular momentum.

The right hand panel of Figure~\ref{fig:vphi_evol} shows that the
rotation speed of enclosed mass shells increases quickly over the
simulated time period which for the $0.1 \, \Msol$ mass shell doubles
in spin from 1/3 to roughly 2/3 of its Kepler value.  This reflects the
growth in mass of the central protostellar core at the expense of disk
material, over this time period.

\bef
\showtwo{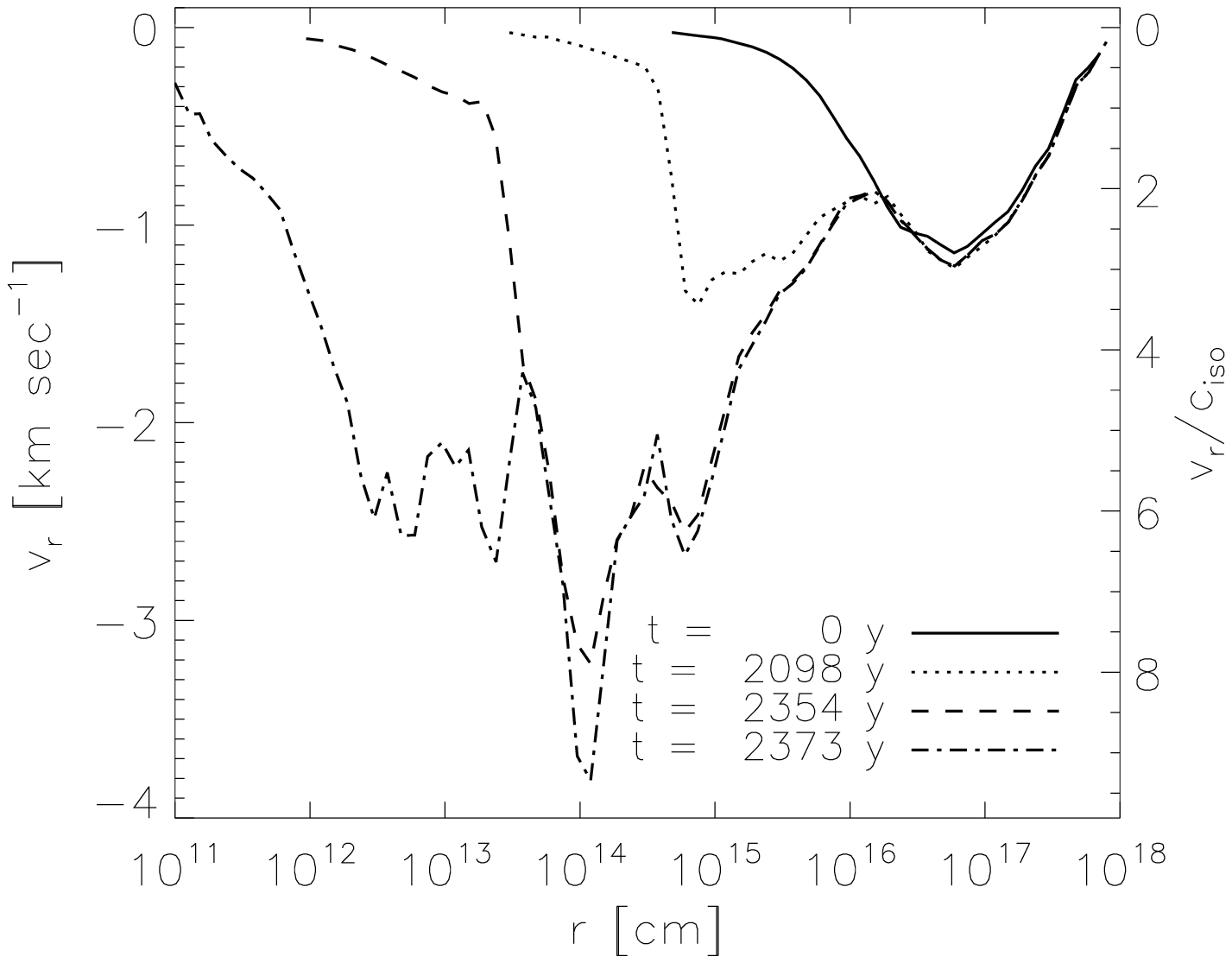}
        {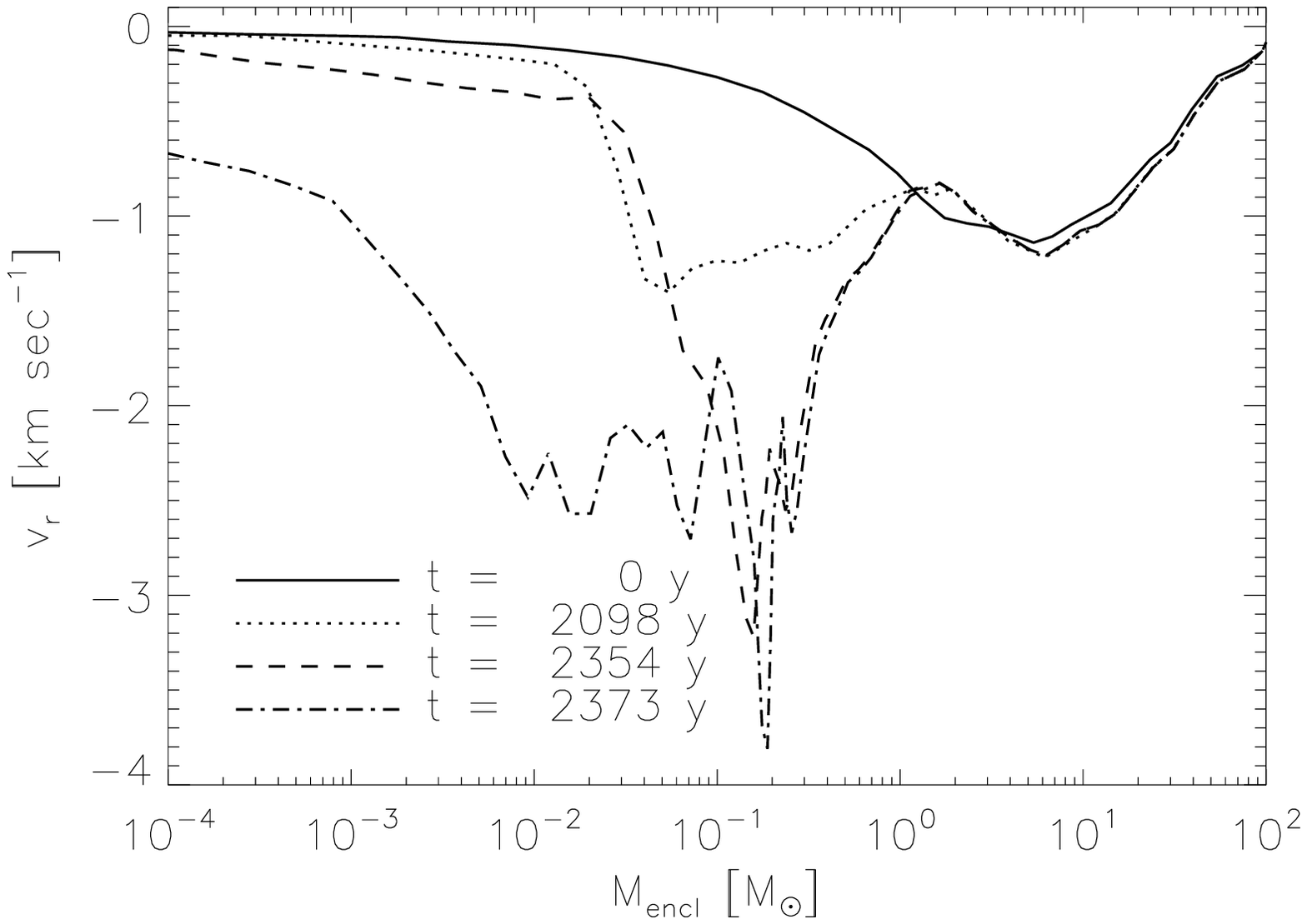}
\caption{The evolution of the radial infall velocity as a function of
  radius (left panel) and enclosed mass (right panel). The collapse
  proceeds from outside-in resembling a collapsing Bonner-Ebert
  sphere. The $t = 0 \, \ys$ graph shows the profile at the beginning
  of our simulation run and the subsequent labels indicate the time
  past since then.}
\label{fig:vrad_evol}
\eef

\section{Massive Star Formation}
\label{sec:starformation}

Although we stopped our simulation at a very early stage into the
formation of the central star (pre-class 0 object) we speculate that
this object will grow quickly to form a massive star since it is
embedded in a massive envelope and the mass accretion is sufficiently
high to overcome the radiation pressure from the burning young massive
star. Most of the material that is destined to form the massive star
in this simulation is assembled by channelling the material flowing
along the spinning filament, through the accretion shock and into the
star by accreting through the disk.  Of greatest importance is the
question of just how large the disk accretion rate is in these early
phases.

We plot the spherically averaged, radial infall velocity as both a
function of radius, as well as of enclosed mass, in
Figure~\ref{fig:vrad_evol} (left and right panels respectively). The
left hand panel shows some of the characteristics of outside-in
collapse that distinguish the collapse of Bonner-Ebert spheres.  In
the left hand panel of this Figure, we see that the core is already in
a state of infall in the initial state of our simulation. The outer
envelope has begun to collapse while most of the interior material is
still nearly at rest.  As the collapse develops, the core radius
progressively shrinks as its density grows.  At the same time, the
maximum infall velocity continues to grow. The outside-in velocity
structure is essentially the same as that seen in the collapse of
Bonner-Ebert spheres, wherein the ever increasing density of the core
is supplied by collapsing material in the envelop.

\bef
\showtwo{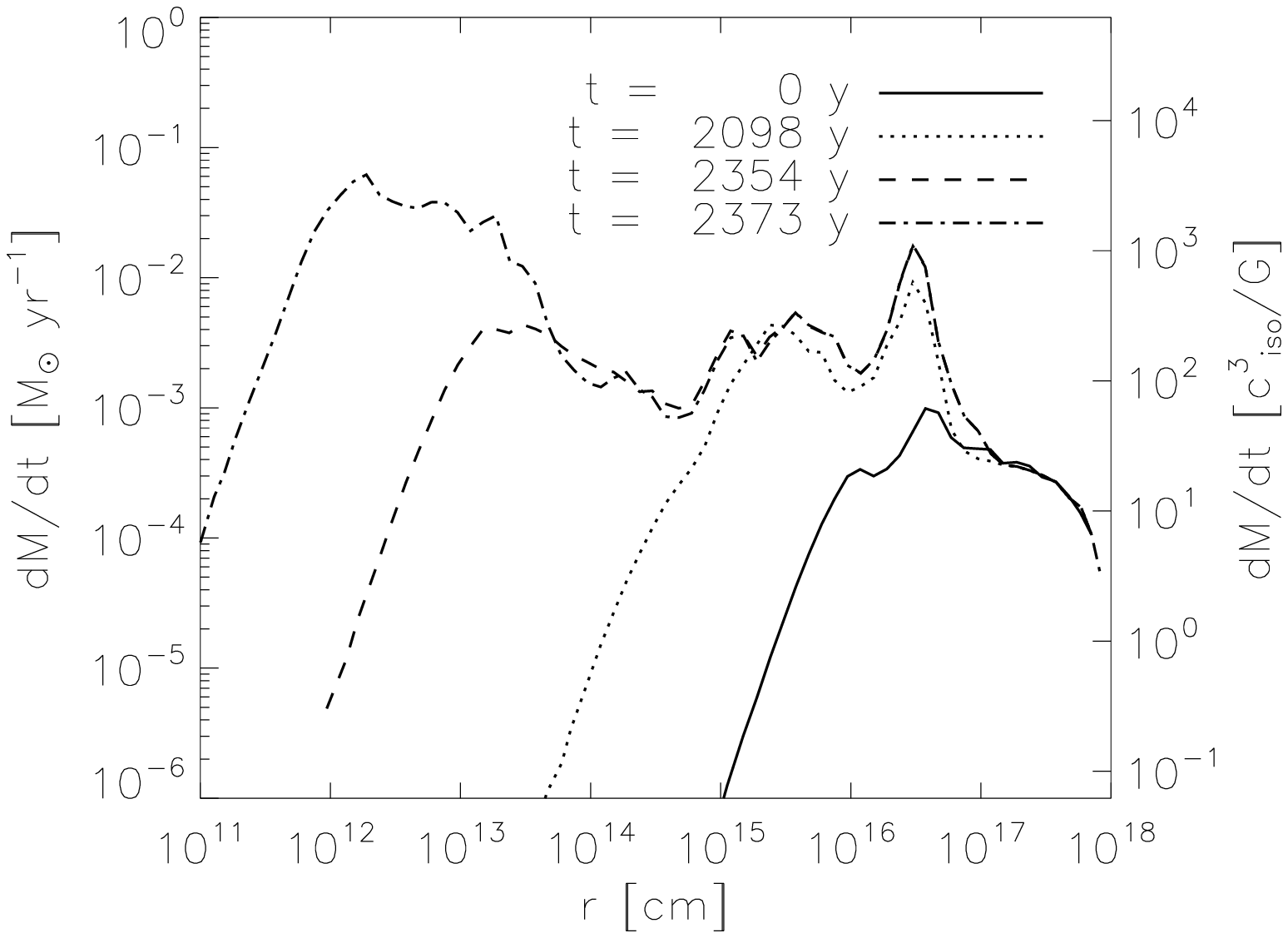}
        {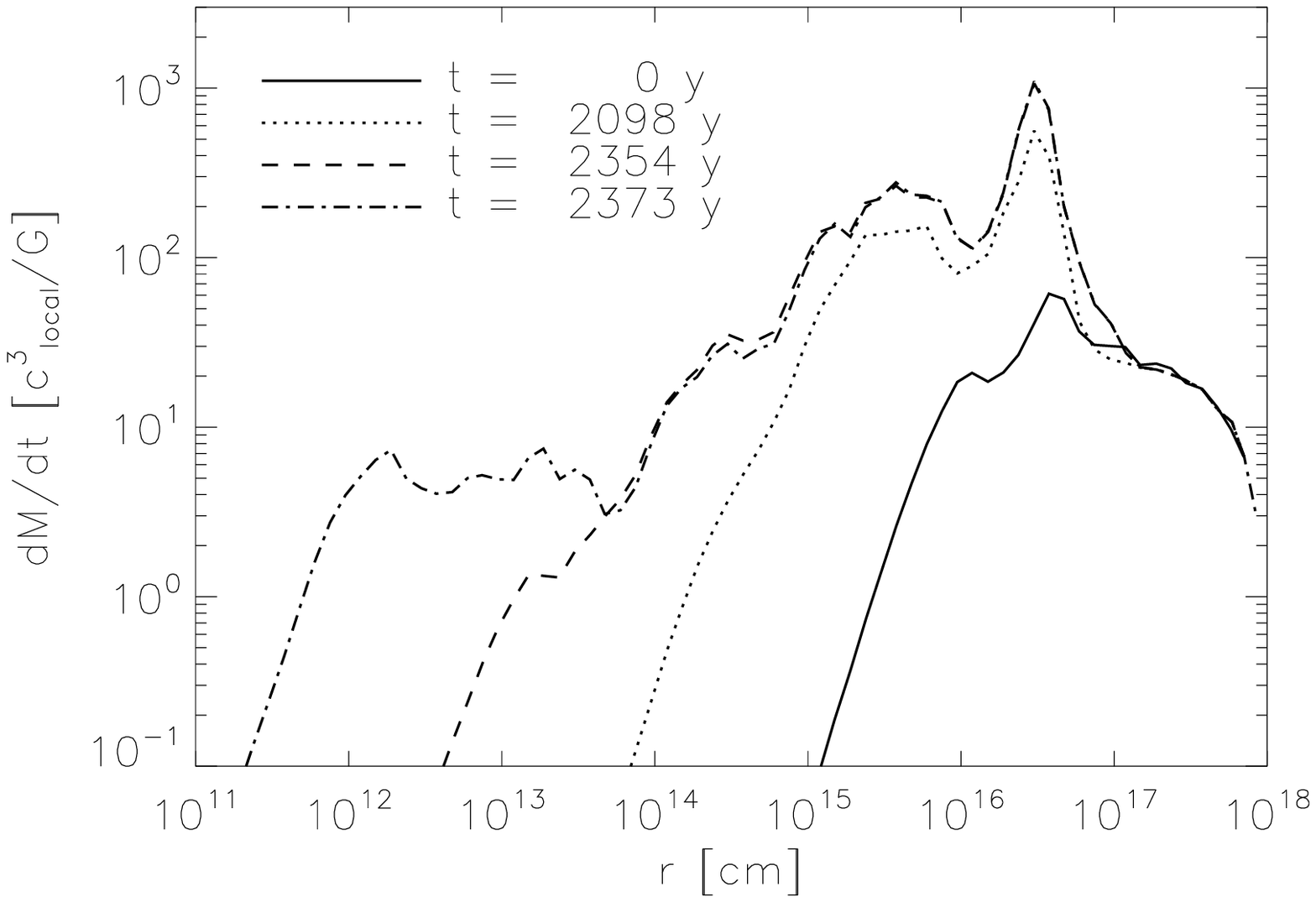}
\caption{The evolution of the mass accretion through spherical shells
  around the core center. The magnitude of the mass accretion is shown
  in absolute values and scaled to the isothermal quantity
  $c_{\sm{iso}}^3/G$ (left panel), and scaled to the {\em local}
  $c_{\sm{local}}^3/G$ accretion measure (right panel). The $t = 0
  \, \ys$ graph shows the profile at the beginning of our simulation
  run and the subsequent labels indicate the time past since then.}
\label{fig:mdot_evol}
\eef

The collapsing gas eventually reaches a density, on small enough
scales, that cooling is too slow and a shock is formed as the outer
material arrives at the position of the more slowly contracting denser
material.  This happens first at a radial scale of a few $10^{14} \,
\cm$.  This is the well known appearance of the first core, whose 
interior contracts almost adiabatically \citep[e.g.][]{Larson69}.  The
2D spatial cuts shown in Figure~\ref{fig:temp_slices_z} clearly show
that this is indeed the scale of the first shock, where dust cooling
in the optical thick regime has become too inefficient to prevent the
appearance of a shock wave. This change in the cooling behavior of
the gas also appears as the change in density behavior seen in the
left panel of Figure~\ref{fig:dens_evol}.  This scale can be predicted
theoretically from the cooling function itself, as we have noted in
previous papers \citep[see e.g.][]{Banerjee06}.

Another interesting aspect of the collapse is that the infall speed is
much higher than the original isothermal sound speed that
characterizes the initial state, and reaches values of Mach 6 on
$10^{15} \, \cm$ scales.  It is also higher than  the asymptotic
infall velocity of collapsing, isothermal Bonner-Ebert sphere where
the Mach number reaches $\sim 3$. The reason for the exceedingly high
infall rate in our case is the initial supersonic turbulence. 
The right hand
panel of Figure~\ref{fig:vrad_evol} shows the infall speed as a
function of the mass enclosed. Only a small fraction of the infalling
material reaches such high Mach number, typically less than $10 \%$ at
these early times.

Perhaps the most important aspect of filamentary accretion is in
driving high accretion rates of material onto the disk.
We show the spherically averaged accretion rates as a
function of disk radius in the left panel of
Figure~\ref{fig:mdot_evol}, in both absolute terms, and then scaled
with respect to the initial isothermal sound speed, $c_{\sm{iso}}$, of
the gas at the beginning of the simulation.  As expected from the
velocity behavior of the collapse shown in the previous
Figure~\ref{fig:vrad_evol}, the accretion rate peaks at some radius,
and this peak moves inwards with time to smaller and smaller scales.
The initial state is already collapsing, as already noted and expected
from the TP04 simulation. The peak of the mass accretion
rate is located at the same radius at which the infall speed achieves
its highest value (compare left hand panels of
Figs.~\ref{fig:vrad_evol} and \ref{fig:mdot_evol}).  Note that at $1
\, \AU$ the infall rate is several $10^{-3} \, \Msol \, \yr^{-1}$
which in the last frame is 10 times higher on sub $\AU$ scales.  In
terms of the initial isothermal sound speed of the simulation, the
infall accretion rate has reached more than $10^3 c^3/G$!  This value
is about 1000 times higher than predicted by the self-similar solution
of singular collapse \citep{Shu77} and about 20 times higher than one
would expect for Bonner-Ebert collapse, as first calculated
by~\citet{Penston69} and \citet{Larson69} \citep[see
also][]{Hunter77}.

The right hand panel of Figure~\ref{fig:mdot_evol} once again plots
the accretion rate as a function of disk radius, except normalized
with respect to {\em the local} sound speed, $c_{\sm{local}}^3/G$.
This latter quantity is certainly not isothermal, and is the sound
speed in the local gas that is undergoing the many cooling processes
that we discuss in the appendix. This local sound speed is of course
greater than the initial isothermal sound speed that we start the
simulation with because the gas heats substantially as a consequence
of shocks.  Thus, as the temperature rises towards the disk's
interior the local sound speed increases as does the value of
$c_{\sm{local}}^3/G$.  In the last time plotted, the ratio of the
infall rate to that computed from this local sound speed is only a
factor of 10 larger at scales around one AU.

\section{Discussion and Conclusions}
\label{sec:discussion}

Our high resolution simulations of a cluster forming molecular clumps
show that the formation of large scale filaments plays a major role in
the formation of accretion disks and massive stars.  Our AMR technique
tracked the collapse of the most massive structure which was the first
to form in the cluster-forming clump which is our initial state.  The
rapid assembly of a massive star - at accretion rates
that are an order of magnitude greater than analytic estimates - 
arises for 2 basic reasons: (i)
supersonic turbulence rapidly sweeps up dense, large-scale structures
by compressing a large volume of gas into sheets and filaments, and
(ii) the high accretion rates in filaments that features
the flow of some material at unusually high Mach numbers. 

\subsection{Collapse in filaments}

The cooling function for gas in the initial state dictates that the
dusty molecular gas remains essentially isothermal until densities of
the order $10^{10} - 10^{11} \, \cm^{-3}$.  The shock waves enormously
compress the material, by factors of order $\Ma^2$ where $\Ma$ is the
Mach number.  For our simulation, the initial rms Mach number
(characterizing the entire turbulent spectrum) is 5, so that the
forming molecular core in this filament has already had an enormous
boost in its background density, and therefore background pressure.

The increase in gas density brought about by the shock then increases
the infall rate in the filament by a large factor.  Since the infall
speed scales as $v_{\sm{ff}} \propto \rho^{1/2}$, we anticipate that
gravitational collapse in such a filament is boosted by a factor of $
v_{\sm{ff}} \propto \Ma$ over typical "virial theorem" estimates.

We have also observed that the infall occurs in a fashion somewhat
like the outside-in collapse of a Bonner-Ebert sphere.  It is
important to remember that a B-E sphere collapse has an infall rate
that is much larger than given only by the sound speed; one finds
$\dot{M} \simeq f \, c^3/G$ where $f\simeq 20-50$ as was restated by
\citet{Hunter77}.  We argue that the factor should scale as $f \simeq
\Ma^3$, i.e., that one replaces the sound speed in the virial formula
with the free-fall speed.  Thus, since the detailed numerics show that
a maximum infall speed of $\Ma \simeq 6$ is reached, then $f \simeq
216$.  This value gets us into the right regime but is still not the
factor of $10^3$ enhancement that we see in the most extreme case in
our.  The difference could be due to a combination of factors including
a geometric factor.

The point is that the filamentary collapse delivers an accretion rate
that is greatly in excess of what is needed to create a massive star
in the conditions, on time scales of only a few thousand years from
the point where these filamentary initial conditions have been
created.

\subsection{Role of cooling}

Cooling plays an important role in these processes, because it
ultimately controls how dense structures can become.  As long as the
gas is isothermal, it can become highly compressed as a consequence of
shock waves.  This early isothermal phase lasts until the gas starts
to become optically thick to emission from dust, at densities
exceeding $10^{11} \, \cm^{-3}$.  At this density, gas becomes
optically thick to dust cooling and begins to contract
quasi-adiabatically. This scale is characteristic of the first shock
that appears at $10 - 20 \, \AU$ (see Figure~\ref{fig:temp_slices_z})
that defines the disk envelope.

The second density of importance is at $10^{16} \, \cm^{-3}$ at which
molecular hydrogen begins to dissociate.  We find that an interesting,
self-regulated state arises in which the gas is kept just cool enough
to avoid both the total dissociation of molecular hydrogen, as well as
the preservation of dust grains from evaporating.  Molecular hydrogen
cooling regulates the gas temperature.  The dissociation of molecular
hydrogen is important because it allows the collapse to resume, as has
been noted by many authors \citep[e.g.][]{Larson69}.

\subsection{Conclusions}

Our basic conclusions are:

\begin{itemize}
\item The assembly of protodisks and protostars is a rapid process in
  a supersonic, turbulent environment. 
\item The most massive object forms first within the molecular
  cloud was the first to violate the 
  Truelove condition in our simulation.  
  This raises the question of how feedback from this (massive)
  star will affect the collapse and star formation in the 
   many other smaller cores that formed in our simulation. 
\item The collapse proceeds from outside-in where the accretion
  through the large scale structure (a filament in our case) ``feeds''
  the dense protostellar structure within it.
\item The initial spin of the filament from oblique shocks lead to
  disk formation because of efficient dust cooling until densities of
  $\sim 10^{11} \, \cm^{-3}$.
\item Even in the absence of magnetic fields and/or outflows angular
  momentum is transfered very efficiently through spiral waves
  in the disk.
\item We observe very high accretion rates ($\dot{M} \sim 10^{-2} \,
  \Msol \, \yr^{-1}$) due to filamentary accretion of the supersonic
  gas onto the forming disks.  These rates are $10^3$ times larger
  than predicted by the collapse of singular isothermal spheres.  We
  find that a reasonable scaling for filamentary accretion is $\dot{M}
  \simeq f \, c^3/G$ where the prefactor should scale as $f
  \simeq \Ma^3$, i.e., that one replaces the sound speed in the virial
  formula with the free-fall speed.  Even this does not quite account
  for the full effect, and there is an additional factor of
  order a few that likely is geometric in nature and accounts for
  filamentary, rather than spherical infall geometry.
 
\end{itemize}

\section*{Acknowledgement}
We thank Debra Shepherd and James Wadsley for valuable comments on our
manuscript. RB thanks Simon Glover for instructive discussions
on micro-physical processes. The FLASH code was in part developed by
the DOE-supported Alliances Center for Astrophysical Thermonuclear
Flashes (ASCI) at the University of Chicago. Our simulations were
carried out on clusters of the SHARCNET HPC Consortium of
Ontario. R.E.P. is supported by the Natural Sciences and Engineering
Research Council of Canada.

\appendix

\section{Appendix: Cooling}
\label{apx:cooling}

In this appendix we summarize the cooling processes which we included
in our simulation.

\subsection{Molecular cooling} 

As in \citet{Banerjee04} and \citet{Banerjee06} we include cooling by
molecular line emissions calculated by \citet{Neufeld95}. These
processes are most important in the low density/low temperature
regime. In \citet{Banerjee04} we showed that the collapse will proceed
isothermally until $n \sim 10^{7.5} \, \cm^{-3}$ if only molecular
cooling is considered.

\subsection{Dust cooling}

Cooling by gas--dust energy transfer in the optical thin regime. Here,
we follow the treatment by \citet{Goldsmith01} with the gas--dust
energy exchange rate
\be
\Lambda_{\sm{gd}} = 2\times 10^{-33} \, 
  \left(\frac{n(\Htwo)}{\cm^{-3}}\right)^2 \,
  \left(\frac{\Delta T}{\K}\right) \,
  \left(\frac{T_{\sm{gas}}}{10\,\K}\right)^{1/2} \, 
  \ergs \, \cm^{-3} \, \sec^{-1}
\label{eq:gas-dust-transfer}
\ee
where $\Delta T = T_{\sm{gas}} - T_{\sm{dust}}$. This cooling process
is very efficient until the core becomes optically thick.

We calculate the the dust equilibrium temperature at each time step by
solving
\be
  \Gamma_{\sm{cr}} + \Lambda_{\sm{gd}} - \Lambda_{\sm{dust}} = 0
\label{eq:dust_equilibrium}
\ee
to get the self-consistent dust temperature. In the above equation
$\Gamma_{\sm{cr}}$ and $\Lambda_{\sm{dust}}$ are the heating by cosmic
rays and the energy loss by dust (black-body) radiation,
respectively. As long as $\Delta T > 0$ the dust is also heated by the
gas--dust interactions. The heating rate by cosmic rays is
\citet{Goldsmith01}
\be
  \Gamma_{\sm{cr}} = 3.9\times 10^{-28} \, 
  \left(\frac{n(\Htwo)}{\cm^{-3}}\right) \,
  \ergs \, \cm^{-3} \, \sec^{-1}
\ee
where we assume a shielding parameter of $\chi = 10^{-4}$. 

\subsubsection{Optical thin regime}

The dust grains loose their thermal energy by black-body like
radiation:
\be
  \Lambda_{\sm{dust}} = \kappa \, \tilde{\sigma} \, T_{\sm{dust}}^4
\label{eq:dust_radiation}
\ee
where $\kappa$ is the opacity (in $\cm^{-1}$) and $\tilde{\sigma} =
f\,\sigma$ is, up to a factor of order unity, the Stefan-Boltzmann
constant $\sigma$. Using the values from \citet{Goldsmith01}
\bea
  \kappa & = & 3.3\times 10^{-26} \, 
      \left(\frac{n(\Htwo)}{\cm^{-3}}\right) \,
      \left(\frac{T_{\sm{dust}}}{18.24 \, \K}\right)^2 \, \cm^{-1} 
\label{eq:dust_opacity} \\
  \tilde{\sigma} & = & 6.85\times 10^{-5} \, 
      \ergs \, \cm^{-2} \, \sec^{-1} \, \K^{-4} \\
      & = & 1.209 \, \sigma \nonumber
\eea
the dust energy loss is
\be
  \Lambda_{\sm{dust}} = 6.8 \times 10^{-33} \,
      \left(\frac{n(\Htwo)}{\cm^{-3}}\right) \,
      \left(\frac{T_{\sm{dust}}}{1 \, \K}\right)^6 \,
      \ergs \, \cm^{-3} \, \sec^{-1} 
\ee
  
Solving the dust equilibrium equation~\refeq{eq:dust_equilibrium} we
find that the dust temperature, $T_{\sm{dust}}$, approaches the gas
temperature, $T_{\sm{gas}}$ at densities $n(\Htwo) \simge 10^{7-8} \,
\cm^{-3}$ depending on the initial gas
temperature.

When the dust temperature is close to equal to the gas temperature the
effective gas cooling becomes
\be
\Lambda_{\sm{gd}} \rightarrow  
   \Lambda_{\sm{dust}}(T_{\sm{dust}} = T_{\sm{gas}}) 
\ee
where the dust temperature is replaced by the gas temperature. In this
regime on can estimate the effective equation of state (EoS) by
equating the compressional heating with the above cooling 
\be
  \frac{{3\over 2} \, n \, k T}{t_{\sm{ff}}} = 
     \kappa \, \tilde{\sigma} \, T_{\sm{gas}}^4
\label{eq:equilibrium_compr1}
\ee
where $t_{\sm{ff}} = \sqrt{3\pi/32\,G_N\,\rho}$ is the free fall
time and $T_{\sm{dust}}$ is replaced by $T_{\sm{gas}}$ in
$\kappa$. From Eq.~\refeq{eq:equilibrium_compr1} we see that $n
\propto T^{10}$ which results in an effective adiabatic index, i.e.
$\gamma_{\sm{eff}} = 1 + \di\ln T/\di\ln n$, of
\be
  \gamma_{\sm{eff}} = \frac{11}{10} 
  \quad \mbox{when} \quad T_{\sm{dust}} \sim T_{\sm{gas}} \, .
\ee

This result shows that the dust is still a very efficient coolant even
if the dust temperature becomes almost equal to the gas
temperature. We summarize the temperature trajectory and the evolution
of the effective equation of state in the
Figures~\ref{fig:evol_simulation} and \ref{fig:gamma_simulation} which
show the result of one of our non-rotating, non-magnetized, spherical
collapse simulations. The simulation results confirm the existence of
the transition region from the isothermal collapse to
$\gamma_{\sm{eff}} = 1.1$ at densities $n \simge 10^{9} \, \cm^{-3}$.

\bef
\showone{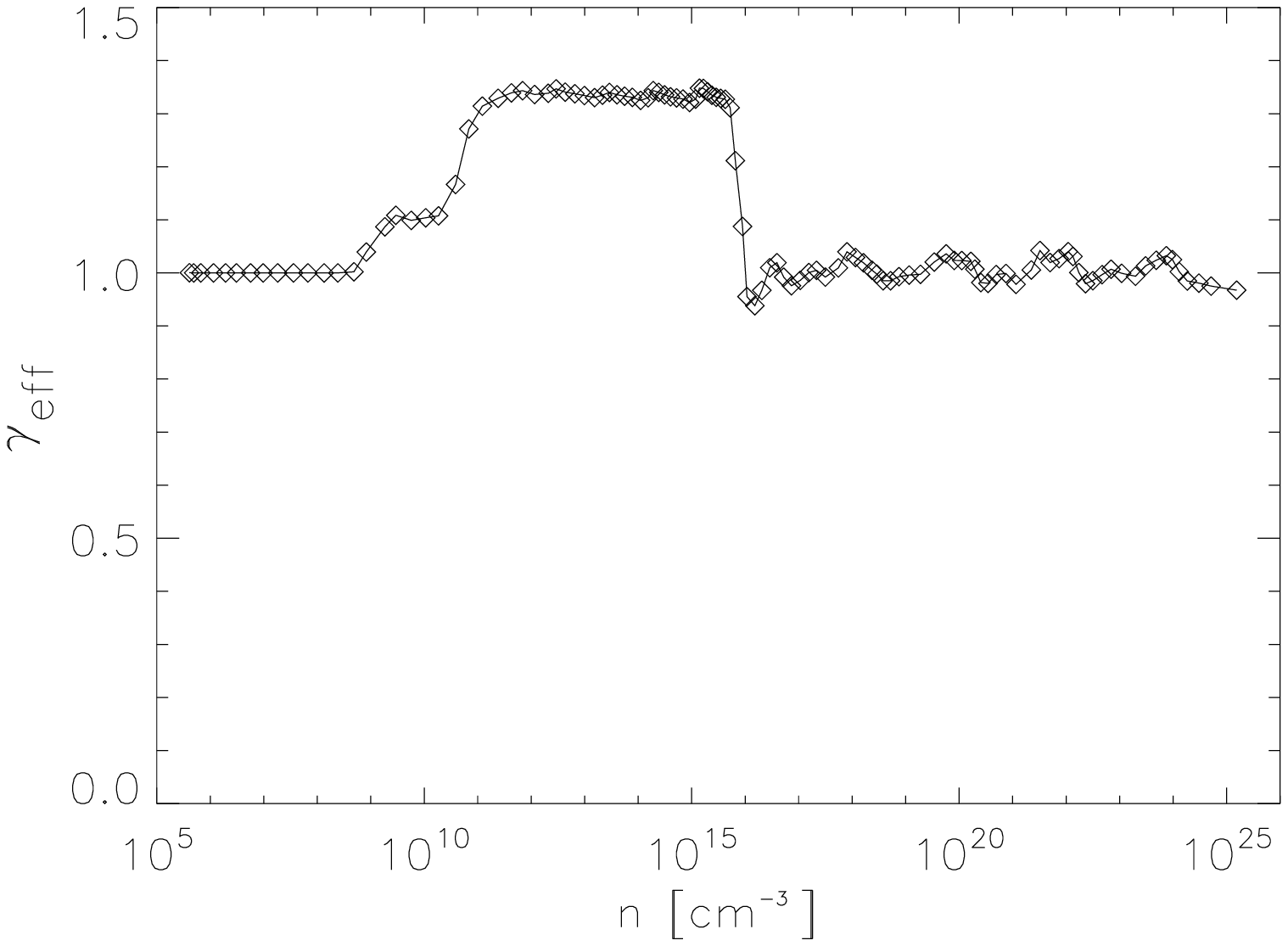}
\caption{Shows the effective EOS as a function of density from our
  non-rotating, non-magnetized spherical, collapse simulation in the case
  where dust cooling and $\Htwo$ dissociation is included. Clearly,
  two distinct transitions can be identified: at a density of $\sim
  10^{11} \, \cm^{-3}$ the core becomes optically thick and contracts
  on quasi-static trajectory, and at $\sim 10^{16} \, \cm^{-3}$ $\Htwo$
  dissociation becomes effective whereby slightly cooling the core region.}
\label{fig:gamma_simulation}
\eef

\subsubsection{Optical thick regime}

We treat the propagation of the radiation field in the optical thick
regime, i.e. $\tau > 1$ (see Sec.~\ref{subsec:cooling} for our
definition of the optical depth $\tau$), as a diffusion process using
the following approximation. The energy loss carried by the radiation
flux ${\bf F}$ is \be \Lambda_{\sm{rad}} = - \nabla\cdot{\bf F} \,.
\label{eq:rad_by_flux}
\ee
In the Eddington approximation the flux can be written as
\citep[e.g.,][]{Turner01} 
\be
  {\bf F} = - \frac{c}{3\, \kappa} \, \nabla \, E
\label{eq:rad_flux}
\ee
were $c$ is the speed of light and $E = a_{\sm{rad}} \,
T_{\sm{dust}}^4$ ($a_{\sm{rad}} = \frac{4\,\sigma}{c} = 7.565\times
10^{-15} \, \ergs \, \cm^{-3} \, \K^{-4}$) is the radiation energy
associated with the dust.

In accordance with the approximation of the optical depth,
Eq.~\refeq{eq:optical_depth}, we use the Jeans length, $\lambda_J$, as
the typical length scale over which the the radiation field changes
significantly. This allows us to approximate $\nabla \approx
f/\lambda_J$ and write Eq.~\refeq{eq:rad_by_flux} as
\footnote{The factor $f$ could be a function of density and
  temperature which can be determined by comparing simulation results
  using the diffusion approach, i.e. Eqs.~\refeq{eq:rad_by_flux} and
  \refeq{eq:rad_flux}, with results using
  Eq.~\refeq{eq:optical_thick_radiation}. Here, we use a constant
  value which we determine by the comparison of
  Eq.~\refeq{eq:time_evolution} with the results of our collapse
  simulation of a non-rotating, non-magnetized Bonnor-Ebert-Sphere.}
\be
  \Lambda_{\sm{rad}} \approx \frac{4}{3} \, f^2 \,
     \frac{\sigma\,T_{\sm{dust}}^4}{\kappa \, \lambda_J^2} 
     \quad \mbox{(optical thick regime)} \, .
\label{eq:optical_thick_radiation}
\ee
We note that the local treatment of the radiation energy loss,
Eq.~\refeq{eq:optical_thick_radiation}, is a good approximation in an
ongoing collapse situation where the Jeans length is the dominant
scale of the system. The approximation becomes less accurate in a
system which is governed by different independent length scales, like
an equilibrium disk with a scale height $h$.

The expression in Eq.~\refeq{eq:optical_thick_radiation} sets an upper
limit of the radiation loss in the optical thick regime and the dust
cooling (Eq.~\refeq{eq:dust_radiation}) is determined by
\be
  \Lambda_{\sm{dust}} = \Lambda_{\sm{rad}}
\ee
which leads to the replacement of the dust opacity in the optical
thick regime 
\be \kappa \rightarrow \tilde{f}/\lambda_J 
  \quad \mbox{(optical thick regime)} \, .  
\ee
where $\tilde{f} = \sqrt{4\sigma/3\tilde{\sigma}}\,f \approx
1.05\,f$.

In summary, we determine the dust temperature in the optical thin {\em
and} optical thick regime by solving Eq.~\refeq{eq:dust_equilibrium}
(using an iterative Newton method) with
\be
  \Lambda_{\sm{dust}} = \kappa_{\sm{eff}} \, \tilde{\sigma} \, 
        T_{\sm{dust}}
\label{eq:dust_radiation2}
\ee
with
\be
  \kappa_{\sm{eff}} \equiv \min\left(\kappa,
  \tilde{f}\,\lambda_J^{-1}\right)
\,.
\ee

The cooling in the optical thick regime is less efficient than in the
optical thin regime. Therefore, the timescale which governs the evolution
of the core becomes larger than the dynamical time scale of the core
region and is given by the cooling time
\bea
t_{\sm{evol}} & = & t_{\sm{cool}} = 
  \frac{{3\over2}\, n \, k T}{\tilde{\sigma} \, T^4 \, \tilde{f} / \lambda_J}
\label{eq:time_evolution} \\
& = & 2.1 \, \tilde{f}^{-1} \, \left(\frac{n(\Htwo)}{10^{10} \, \cm^{-3}}\right)^{1/2} \, 
      \left(\frac{T}{100\, \K}\right)^{-5/2} \, \ys \nonumber
\eea

\bef
\showone{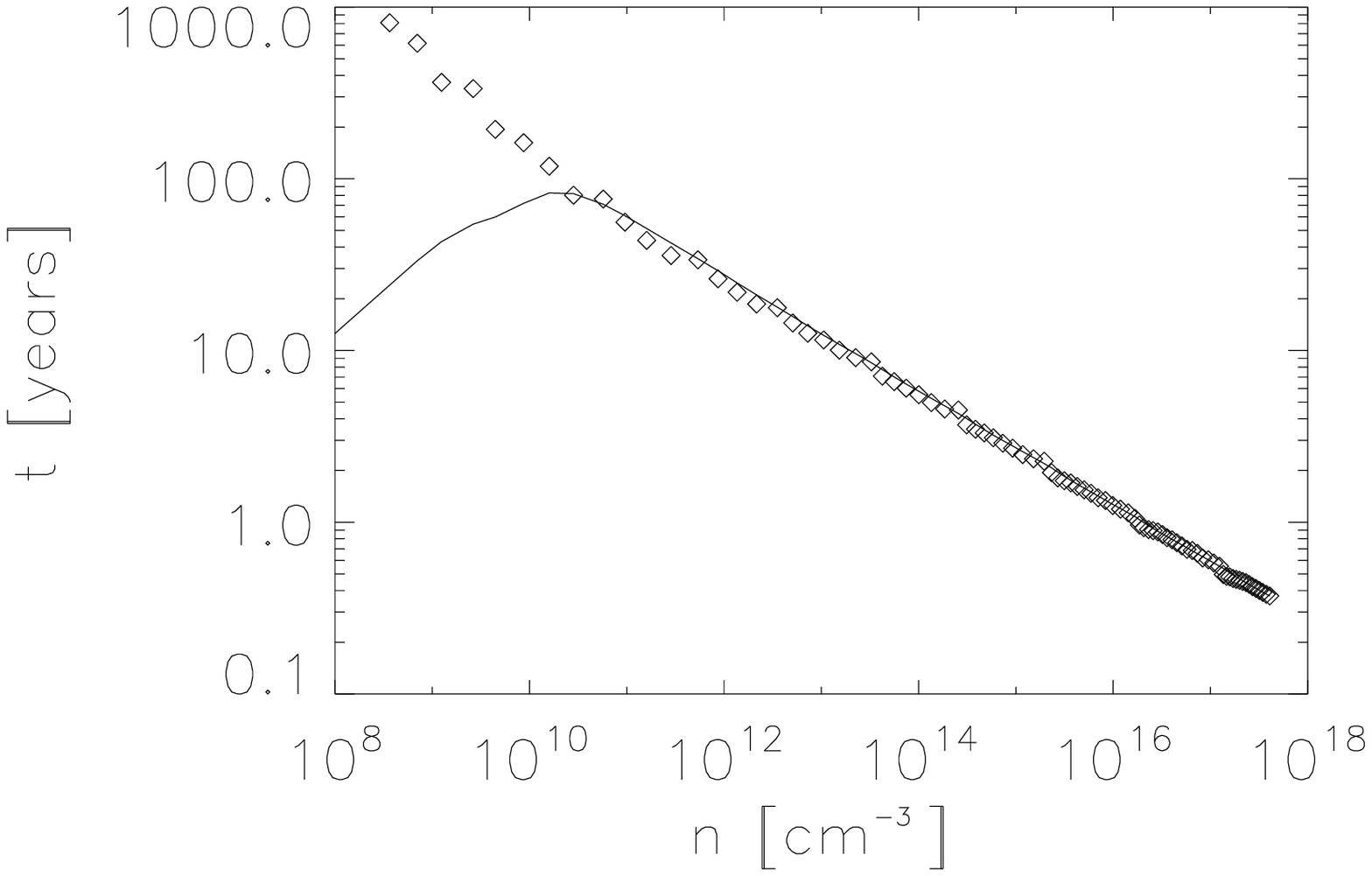}
\caption{Shows the evolution time of the density, i.e. $t_{\sm{evol}}
  = n/\dot{n}$ (diamonds) from our simulation and the time calculated
  from the prediction Eq.~\refeq{eq:time_evolution} (line) where we
  set $\tilde{f} = 1.67$. The simulation follows closely the
  trajectory of our theoretical estimate of
  Eq.~\refeq{eq:time_evolution}}
\label{fig:times_sim}
\eef

In Fig.~\ref{fig:times_sim} we show the density dependence of the core
evolution time, $t_{\sm{evol,sim}} \equiv n/\dot{n}$ from our
simulation and the timescale from Eq.~\refeq{eq:time_evolution}. Both
timescales have the same dependence on density ($\propto
n^{-1/3}$) in the optical thick regime and the same amplitude if we
choose $\tilde{f} = 1.67$.

Equation~\refeq{eq:time_evolution} does not constrain the
density--temperature relation (i.e. the equation of state). But as the
system collapses on a trajectory at which the thermal pressure
is (almost) in balances with gravity this trajectory will be the path at which
\be
\gamma_{\sm{eff}} = \gamma_{\sm{crit}} = 4/3 \, .
\ee

Figs.~\ref{fig:gamma_simulation} and \ref{fig:evol_simulation} show the
evolution of the effective EoS ($\gamma_{\sm{eff}} = \di\ln p / \di\ln
\rho$) and the temperature of the core region of our spherical
collapse simulation. The collapse in the optical thick regime follows
the path where $\gamma_{\sm{eff}} \approx \gamma_{\sm{crit}}$.

\subsection{Warm temperature regime and dust melting}

At temperatures of about $100 \, \K$ the dust opacity becomes almost
independent of the frequency and at temperatures around $1500 \, \K$
the dust opacity decreases sharply due to the melting of dust grains
\citep[for a recent compilation of opacities see][]{Semenov03}. To
incorporate these effects we fit the dust opacity by a piecewise power
law (see also Eq.~\refeq{eq:dust_opacity})
\be
\kappa \propto \left\{
  \begin{array}{rcl}
    T^2      & \quad ; \quad & T < 200 \, \K \\
    T^0      & \quad ; \quad & 200 \, \K  < T < 1500 \, \K \\
    T^{-12}  & \quad ; \quad & T > 1500 \, \K
  \end{array} \right.
\ee

\subsection{$\Htwo$ formation and dissociation}

We use the \citet{Hollenbach79} rates to account for the formation of
$\Htwo$. These authors compute the formation of hydrogen molecule on
dust grain surfaces from collisions with atomic hydrogen. The
formation coefficient, $R_f$, can be written as:
\be
R_f = 3\times 10^{-18} \,
\frac{T^{1/2} \, f_a}{1 + 0.04\,(T + T_{\sm{dust}})^{1/2} +
      2\times 10^{-3}\,T +  8\times 10^{-8} \,T^2} \, ,
\label{eq:h2_formation_coeff}
\ee
where $T$ and $T_{\sm{dust}}$ are given in $\K$ and the factor $f_a$
is given by
\be
f_a = \frac{1}{1 + 10^4 \, \exp{(-600/T)}} \, .
\ee

To account for the process of $\Htwo$ dissociation in the hot dens
region we use the dissociation rates from \citet{Shapiro87} which are
recalculated functions based on the work by \citet{Lepp83}. These
dissociation rates are calculated from the collisional breakup
of molecular hydrogen out of all 15 vibrational levels.

The dissociation rate coefficient can be well fitted to the form
\be
\log{k_D(n,T)} = \log{k_H} - \frac{\log{k_H/k_L}}{1 + n/n_{\sm{tr}}} \, ,
\label{eq:diss_coeff}
\ee
where $k_D$ is given in units $\cm^3 \, \sec^{-1}$ and $k_H$ and $k_L$
are the coefficients for the low and high density regimes,
respectively
\bea
k_L & = & 1.12\times 10^{-10} \, \exp{(-7.035\times 10^4 \,\K/T)} \,
          \cm^3 \, \sec^{-1} \, ;
          \quad \mbox{H -- $\Htwo$} \\
    & = & 1.18\times 10^{-10} \, \exp{(-6.950\times 10^4\,\K/T)} \, 
          \cm^3 \, \sec^{-1} \, ; 
          \quad \mbox{$\Htwo$ -- $\Htwo$} \nonumber \\
k_H & = & 1.20\times 10^{-9} \, \exp{(-5.24\times 10^4\,\K/T)} \,
          \cm^3 \, \sec^{-1} \, ; 
          \quad \mbox{H -- $\Htwo$} \\
    & = & 1.30\times 10^{-9} \, \exp{(-5.33\times 10^4\,\K/T)} \,
          \cm^3 \, \sec^{-1} \, ; 
          \quad \mbox{$\Htwo$ -- $\Htwo$} \, . \nonumber
\eea
The transitional density $n_{\sm{tr}}$ in units of $\cm^{-3}$ is given by
\bea
\log{n_{\sm{tr}}} & = & 4.00 - 0.416\,x - 0.327\,x^2 
  \, ; \quad \mbox{H -- $\Htwo$} \\
  & = & 4.845 - 1.3\,x + 1.62\,x^2
  \, ; \quad \mbox{$\Htwo$ -- $\Htwo$} \nonumber
\eea
with
\be
x = \log{T/10^4 \, \K} \, .
\ee

The binding energy of $\Htwo$ molecules is $4.48\,\mbox{eV}$. For each
collision with molecular hydrogen which leads to its dissociation this 
energy is lost from the (hydrogen) gas. Therefore, the effective
cooling rate due to hydrogen dissociation is given by
\be
\Lambda_{\sm{diss}} = 4.48\,\mbox{eV} \, n \, n(\Htwo) \, k_D \, ,
\label{eq:h2_dissociation}
\ee
where $n$ is either $n(\Htwo)$ or $n(\mbox{H})$ depending on the
collisional partner.

The FLASH code provides data arrays which store the mass, $X_i$,
fraction of different species. We use this ability to keep track of
the number densities of hydrogen in its different forms
(i.e. molecular or atomic). The number density of the species $i$ is
given by
\be n_i = X_i \, \frac{\rho}{\mu_i \, m_H} \, ,
  \label{eq:density_by_fraction}
\ee
where $\mu_i$ is the atomic weight of the species $i$, $\rho$ is the
total matter density, and $m_H$ is the hydrogen mass.

We point out that we did not include effects from $\Htwo$ formation by
three body processes which could still be important in the high
density ($> 10^{16} \, \cm^{-3}$) and warm temperature ($< 1100 \, K$)
regime. We might therefore slightly overestimate the cooling ability
from $\Htwo$ dissociation. Three body formation rates are rather
uncertain and different compilations can be found, for instance, in
\citet{Abel02, Palla83}.

\bibliographystyle{mn2e}
\bibliography{astro}

\end{document}

%% file: journals.tex
\def\jnl@style#1{{\rmfamily#1}}%
\def\jref@jnl#1{{\jnl@style#1}}%

\newcommand\aj{\jref@jnl{AJ}}%
\newcommand\araa{\jref@jnl{ARA\&A}}%
\newcommand\apj{\jref@jnl{ApJ}}%
\newcommand\apjl{\jref@jnl{ApJ}}%
\newcommand\apjs{\jref@jnl{ApJS}}%
\newcommand\ao{\jref@jnl{Appl.~Opt.}}%
\newcommand\apss{\jref@jnl{Ap\&SS}}%
\newcommand\aap{\jref@jnl{A\&A}}%
\newcommand\aapr{\jref@jnl{A\&A~Rev.}}%
\newcommand\aaps{\jref@jnl{A\&AS}}%
\newcommand\azh{\jref@jnl{AZh}}%
\newcommand\baas{\jref@jnl{BAAS}}%
\newcommand\jrasc{\jref@jnl{JRASC}}%
\newcommand\memras{\jref@jnl{MmRAS}}%
\newcommand\mnras{\jref@jnl{MNRAS}}%
\newcommand\pra{\jref@jnl{Phys.~Rev.~A}}%
\newcommand\prb{\jref@jnl{Phys.~Rev.~B}}%
\newcommand\prc{\jref@jnl{Phys.~Rev.~C}}%
\newcommand\prd{\jref@jnl{Phys.~Rev.~D}}%
\newcommand\pre{\jref@jnl{Phys.~Rev.~E}}%
\newcommand\prl{\jref@jnl{Phys.~Rev.~Lett.}}%
\newcommand\pasp{\jref@jnl{PASP}}%
\newcommand\pasj{\jref@jnl{PASJ}}%
\newcommand\qjras{\jref@jnl{QJRAS}}%
\newcommand\skytel{\jref@jnl{S\&T}}%
\newcommand\solphys{\jref@jnl{Sol.~Phys.}}%
\newcommand\sovast{\jref@jnl{Soviet~Ast.}}%
\newcommand\ssr{\jref@jnl{Space~Sci.~Rev.}}%
\newcommand\zap{\jref@jnl{ZAp}}%
\newcommand\nat{\jref@jnl{Nature}}%
\newcommand\iaucirc{\jref@jnl{IAU~Circ.}}%
\newcommand\aplett{\jref@jnl{Astrophys.~Lett.}}%
\newcommand\apspr{\jref@jnl{Astrophys.~Space~Phys.~Res.}}%
\newcommand\bain{\jref@jnl{Bull.~Astron.~Inst.~Netherlands}}%
\newcommand\fcp{\jref@jnl{Fund.~Cosmic~Phys.}}%
\newcommand\gca{\jref@jnl{Geochim.~Cosmochim.~Acta}}%
\newcommand\grl{\jref@jnl{Geophys.~Res.~Lett.}}%
\newcommand\jcp{\jref@jnl{J.~Chem.~Phys.}}%
\newcommand\jgr{\jref@jnl{J.~Geophys.~Res.}}%
\newcommand\jqsrt{\jref@jnl{J.~Quant.~Spec.~Radiat.~Transf.}}%
\newcommand\memsai{\jref@jnl{Mem.~Soc.~Astron.~Italiana}}%
\newcommand\nphysa{\jref@jnl{Nucl.~Phys.~A}}%
\newcommand\physrep{\jref@jnl{Phys.~Rep.}}%
\newcommand\physscr{\jref@jnl{Phys.~Scr}}%
\newcommand\planss{\jref@jnl{Planet.~Space~Sci.}}%
\newcommand\procspie{\jref@jnl{Proc.~SPIE}}%